\documentclass[10pt]{iopart}
\usepackage{graphicx}% Include figure files
\usepackage{iopams}
\usepackage{setstack}
\usepackage{cite}
\begin{document}

\title[Multiple Dirac eikonal $p-A$ scattering]{Multiple Dirac eikonal scattering of polarized intermediate-energy protons by nuclei}

\author{V.V. Pilipenko and V.I. Kuprikov}

\address{National Science Center “Kharkov Institute of Physics and Technology”, Kharkov 61108, Ukraine}
\ead{vpilipenko@kipt.kharkov.ua}
\vspace{10pt}
%\begin{indented}
%\item[]December 2020
%\end{indented}

\begin{abstract}
An improved model of the multiple Dirac eikonal scattering of proton on nucleons of the target nucleus is considered. In this model, the amplitudes of elastic $p-A$ scattering are found on the basis of the Watson series of multiple scattering by means of the eikonal expansion of the Dirac propagator for the free proton motion between scattering events on nucleons, and the nucleus structure is described in the relativistic mean field model. The calculations have been performed for the complete set of observables for the elastic $p+^{40}$Ca and $p+^{208}$Pb scattering at 800 MeV. The effects of allowing for distinction between the relativistic scalar and vector nucleon densities on the description of the $p-A$ scattering observables have been studied, as well as the results of calculations using nucleon densities obtained in different modern variants of the relativistic mean field model have been compared.
\end{abstract}

\vspace{2pc}
\noindent{\it Keywords}: Proton–nucleus scattering, polarization observables, multiple scattering, Dirac equation, eikonal approximation
%\maketitle

\section{Introduction}
At present, the theoretical analysis of experimental data on the processes of proton-nucleus ($p-A$) interaction in the intermediate energy region (of the order of several hundred MeV) continues to be an important task, because in this way one may hope to derive valuable information both on nuclear reaction mechanisms and about the structure of nuclei, in particular, about distribution densities of nucleons in nuclei and distinctions between the proton and neutron densities, as well as about nucleon correlations. The most commonly employed approaches to describing these processes are the multiple diffraction scattering theory (MDST) and the relativistic impulse approximation (RIA). In the MDST, which was first proposed by Glauber in \cite{Gla1} and further developed in a great number of works (see, e.g., \cite{Gla2,Har,Alk,Osl,Wall1,Wall2,Ble1,Kup1,Kup2,Ber} and references therein), the amplitude of the $p-A$ interaction is expressed in terms of the elementary nucleon-nucleon ($NN$) scattering amplitudes and the target-nucleus wave function (primarily the nucleon distribution densities) within the eikonal approximation. The RIA approach (see \cite{McN,Ray2,Mur,Tjon,Wall3,Ray,Clar,Sak,Kaki,Yah} and references therein) is based on solving the Dirac equation with a relativistic microscopic optical potential, which is also constructed from the $NN$-amplitudes and the target-nucleus nucleon densities, normally the relativistic ones.

Along with the differential cross sections of the $p-A$ scattering processes, of special interest is analyzing the spin observables, because they are rather sensitive to the choice of the used theoretical models. In the case of elastic $p-A$ scattering, these are, for example, the analyzing power and spin rotation function, which form, together with the cross section, the complete set of observable quantities for the description of this process on spin-zero nuclei. For purposes of this kind of investigations, sufficiently accurate theoretical models are required for reliably analyzing such data. At present, an extensive information from the phase analysis of $NN$-scattering is available, which has significantly improved the knowledge of $NN$-amplitudes, including their dependence on the spins of nucleons, needed in the MDST and RIA approaches. The current state of the art in the development of theoretical studies of the structure of nuclei based on the nonrelativistic Hartree-Fock model with effective Skyrme forces and on the models of the relativistic mean field (RMF, see \cite{Wale,Hor,Gam,Ser,Meng,Typ,Lal}) makes it possible to obtain realistic nuclear densities, which may be used and tested in the MDST and RIA calculations.

As for the RIA model, it has clearly shown its advantages over the nonrelativistic impulse approximation (see, for example, \cite{Wall3,Ray,Sak}) and turned out to be successful in describing the experimental data on cross sections and spin observables for the proton scattering by various stable nuclei in the energy range from lower values about 200 MeV up to higher energies where the Glauber MDST is usually applied, for which a number of precise measurements are available in the literature. Besides the consideration of stable target nuclei, recently, the interest has been rising to employing the RIA approach to the theoretical studies of proton scattering by various unstable nuclei (see, \cite{Sak,Kaki,Yah}  and references therein), owing to the expected future experimental possibilities in this area basing on using radioactive beams.

The MDST has been a widely used approach to analyzing the scattering of protons and other hadrons on atomic nuclei over a long period of time and it is still employed to study various processes induced by the nucleon-nucleus and nucleus-nucleus interaction at intermediate energies. Application of the MDST has been usually yielding reasonable results in describing corresponding experimental data. However, the calculations of complete set of the $p-A$ scattering observables by the MDST with using the above-mentioned realistic $NN$ amplitudes and nucleon densities without involving any free parameters show certain discrepancies between the theoretical results and experimental data (see, e.g.,~\cite{Kup1,Kup2}). Therefore, the MDST in its traditional formulation may be considered as a good first approximation for describing the $p-A$ scattering, which should be refined, in order to perform a more accurate quantitative analysis of the data. Note that different corrections to the original MDST formulation were studied and discussed (see \cite{Har,Alk,Osl,Wall1,Wall2,Ble1} and references therein), which revealed that many of these corrections tend to be mutually compensated, but nevertheless some of them can prove to be significant. In \cite{Kup1,Kup2,Ber}, basing on the realistic $NN$-amplitudes and nuclear densities, we performed the MDST calculations of the $p-A$ scattering observables by the model with allowance for the two-nucleon correlations through including intermediate excitations of target nuclei and also taking into account noneikonal corrections, analogous to those considered in \cite{Ble1}. Although these corrections allowed us to somewhat improve agreement with the experimental data, nevertheless there remained some shortcomings in describing the considered spin observables, which suggests that further refinements of the model should be sought.

In this connection, we should recall that the theoretical basis of MDST calculations of the $pA$ scattering amplitude is in fact nonrelativistic, because they do not describe the relativistic behavior of the spin, and do not use a consistent relativistic description of the nuclear structure, while involve only baryon densities of nucleons. Therefore, it seems advisable to develop a model similar to the MDST, which would be based, as RIA, on the Dirac equation with a simultaneous relativistic description of the target-nucleus structure. Such a model could be a useful complement to the RIA approach for analyzing the experimental data in the energy region of the Glauber MDST applicability (higher than about 500 MeV). Here, we should mention the relativistic multiple-scattering Glauber approximation (see, e.g., \cite{Ryck,Col} and references therein), which was applied in studies of nucleon knockout reactions induced by electrons and protons and processes of photo- and electroproduction of mesons on nuclei. In this model, the involved hadrons and the target nuclei are described by relativistic wave functions and the initial- and final-state interaction of these hadrons with the target-nucleus nucleons is taken into account by means of the standard Glauber multiple scattering profile function, which however does not include the spin dependence.

We suppose that the desired relativistic refinement of the MDST approach can be implemented by considering a more consistent relativistic description of the process of multiple scattering of an incident proton by a system of nucleons that make up the nucleus, which may be performed by means of the high-energy eikonal expansion, analogously to building the MDST as a multiple eikonal scattering approach (see, \cite{Wall1,Wall2,Ble1}), but basing on the Dirac equation. Earlier, we developed an initial version of the multiple Dirac eikonal scattering (MDES) model \cite{Pil}, however, there the sought $p-A$ scattering amplitude was constructed in the simplest form, which was aimed to be most similar to the usual MDST approach. In particular, we used only the baryon densities for describing the nucleus structure and, therefore, possible effects of the difference between the scalar and vector nucleon distribution densities, calculated in the relativistic mean field model, were discarded. A number of other simplifications were also made.

In this paper, we try to provide a more accurate and consistent relativistic formulation of the MDES model of the elastic $p-A$ scattering, in particular, taking into account the differences between the relativistic scalar and vector nuclear densities calculated in the RMF approach. We study the influence of the allowance for the difference of these densities on the description of the complete set of the elastic $p-A$ scattering observables. We also compare the results of MDES calculations of the scattering observables with using nucleon densities obtained in different variants of the RMF model.

\section{Formulation of the theoretical model}
We develop our model employing the high-energy eikonal expansion basing on the Dirac equation. To implement this, we use the method similar to that of \cite{Ble1}, in which the Glauber–Sitenko approach is considered as a theory of multiple eikonal scattering on the target nucleons with making use of the mutual compensation of a number of corrections, according to the results of \cite{Wall1,Wall2}, which allows the following simplifications: rescatterings on the same nucleon and the movement of target nucleons during the $p-A$ interaction process may be neglected; the most essential matrix elements of the $pN$ $t-$operator can be taken as local and determined on the energy surface. Thus, to construct the $T-$matrix operator for the $p-A$ scattering, we proceed from the Watson series of multiple scattering \cite{Gold}, restricting ourselves up to the terms of order $A$ (we assume $\hbar  = c = 1$):
%\begin{equation}
\begin{eqnarray}
T &=& \sum\limits_{j = 1}^A {{t_j}}  + \sum\limits_{j = 1}^A {\sum\limits_{k \ne j}^A {{t_j}{{\tilde G}^{( + )}}{t_k}} }  + \sum\limits_{j = 1}^A {\sum\limits_{k \ne j}^A {\sum\limits_{l \ne k \ne j}^A {{t_j}{{\tilde G}^{( + )}}{t_k}{{\tilde G}^{( + )}}{t_l}} }} \nonumber \\
&&+ ... + \mbox{term\;of\;order\;{\it A}},
\label{Eq1}
\end{eqnarray}
%\end{equation}
%(1)
\begin{equation}
\eqalign
{{\tilde G^{( + )}} = {\left[ {E - {H_p}\left( {{\bf{\hat k}}} \right) - {H_t}\left( {{\bf{\hat k}},\left\{ {{{\bf{r}}_j}} \right\}} \right) + i0} \right]^{ - 1}}, \; \\
{H_p}\left( {{\bf{\hat k}}} \right) = {\balpha \bf{\hat k}} + \beta m, \;
{H_t}\left( {{\bf{\hat k}},\left\{ {{{\bf{r}}_j}} \right\}} \right) = \sqrt {{{{\bf{\hat k}}}^2} + {h_t}^2\left( {\left\{ {{x_j}} \right\}} \right)} ,}
\label{Eq2}
\end{equation}
%(2)
where ${t_j}$ are the proton-nucleon scattering operators, which are taken in the $p-A$ center-of-mass (c.m.) frame on the energy surface. The propagator of the free motion of the proton ${\tilde G^{( + )}}$ contains the Dirac Hamiltonian ${H_p}$ of the proton and the Hamiltonian of the spin-zero target nucleus ${H_t}$, which has a model form \cite{Rem} providing the correct relativistic kinematic relations, where ${h_t}\left( {\left\{ {{x_j}} \right\}} \right)$ is the Hamiltonian of the internal motion of the nucleus (${x_j}$ are the nucleon variables); ${\bf{\hat k}} =  - i{\partial \left/ \right. {\partial {\bf{r}}}}$ is the momentum operator, ${\bf{r}} = {{\bf{r}}_p} - {{\bf{r}}_t}$, ${{\bf{r}}_p}$ and ${{\bf{r}}_t}$ are the coordinates of the proton and the nucleus center of mass; $E = {\varepsilon _p} + {\varepsilon _t}$ is the energy of the system, ${\varepsilon _p} = \sqrt {k_0^2 + {m^2}} $ and ${\varepsilon _t} = \sqrt {k_0^2 + {M^2}} $ are the energies of the proton and nucleus, and $m$ and $M$ are their masses, ${k_0} = \left| {{{\bf{k}}_i}} \right| = \left| {{{\bf{k}}_f}} \right|$ is the magnitude of the initial and final momenta in the $p-A$ c.m. frame.

The elastic $p-A$ scattering amplitude in the $p-A$ c.m. frame is related to the $T-$operator (\ref{Eq1}) as follows:
\begin{equation}
F\left( {{{\bf{k}}_f},{{\bf{k}}_i}} \right) =  - \frac{1}{{4\pi }}\frac{{{\varepsilon _t}}}{{{\varepsilon _p} + {\varepsilon _t}}}\left( {{{\bar u}^{\mu '}}\left( {{{\bf{k}}_f}} \right){\gamma ^0}\Psi _0^ + \left\langle {{{\bf{k}}_f}} \right|T\left| {{{\bf{k}}_i}} \right\rangle {\Psi _0}{u^\mu }\left( {{{\bf{k}}_i}} \right)} \right) ,
\label{Eq3}
\end{equation}
%(3)
where ${\Psi _0}$ is the wave function of the nucleus ground state; $\left| {{{\bf{k}}_{i,f}}} \right\rangle $ are the initial and final plane waves normalized to the unit density; ${u^\mu }\left( {\bf{k}} \right)$ are the proton bispinors with the normalization: ${\bar u^\mu }\left( {\bf{k}} \right) \cdot {u^\mu }\left( {\bf{k}} \right) = 2m$.

The matrix elements (m.e.) $\left\langle {{\bf{k'}}} \right|{t_j}\left( {{{\bf{r}}_j}} \right)\left| {\bf{k}} \right\rangle = \exp \left( {i{\bf{q}}{{\bf{r}}_j}} \right)\left\langle {{\bf{k'}}} \right|{t^{(j)}}\left| {\bf{k}} \right\rangle $ in (\ref{Eq1})  (here we take into account the offset from the nucleus c.m. by the vector ${{\bf{r}}_j}$ as compared to the operator ${t^{(j)}}$ for a nucleon of sort $j$ in the center of coordinates) should be related to the $pN-$scattering amplitude determined in the $pN$ c.m. frame. We may assume that the main role is played by the on-shell m.e. values of the ${t^{(j)}}$ operator, which are considered as local and dependent only on the momentum transfer ${\bf{q}} = {\bf{k}} - {\bf{k'}}$ and the energy value, and the approximation determined for the on-shell m.e. in the $pN$ c.m. frame may be extrapolated for the off-shell m.e. values. The noncovariant $pN$ amplitude ${f_j}\left( {{{\bf{q}}_c}} \right)$ in the $pN$ c.m. frame has the following form (${{\bf{q}}_c} = {{\bf{k}}_c} - {{\bf{k'}}_c}$ is the corresponding momentum transfer):
%\begin{equation}
\begin{eqnarray}
{f_j}\left( {{{\bf{q}}_c}} \right)& =& {A_j}\left( {{q_c}} \right) + {B_j}\left( {{q_c}} \right){{\bsigma}_p}{{\bsigma}_j} + {C_j}\left( {{q_c}} \right)\left( {{{\bsigma}_p} + {{\bsigma}_j}} \right){\bf{n}} \nonumber \\
&&+ {D_j}\left( {{q_c}} \right)\left( {{{\bsigma }_p}{{\bf{q}}_c}} \right)\left( {{{\bsigma }_j}{{\bf{q}}_c}} \right) + {E_j}\left( {{q_c}} \right){\sigma _{pz}}{\sigma _{jz}},
\label{Eq4}
\end{eqnarray}
%\end{equation}
%(4)
where ${{\bsigma }_p}$ and ${{\bsigma }_j}$ are the Pauli matrices of the proton and nucleon, the average momentum direction is ${{\bf{e}}_z}\parallel {{\bf{k}}_{ca}} = ({{\bf{k}}_c} + {{\bf{k'}}_c})/2$ and ${{\bf{n}}_c} = {{\bf{\hat q}}_c} \times {{\bf{e}}_z}$. The noncovariant m.e. of ${t^{(j)}}$ is related to the amplitude ${f_j}\left( {{{\bf{q}}_c}} \right)$ as follows: $\left\langle {{{{\bf{k'}}}_c}} \right|{t^{(j)}}{\left| {{{\bf{k}}_c}} \right\rangle _{\rm{nc}}} =  - 4\pi {f_j}\left( {{{\bf{q}}_c}} \right)/{\varepsilon _{pc}}$. Further, we should introduce the Lorentz-invariant $pN$ amplitude, which may be represented in the form (see, e.g., \cite{McN}):
%\begin{equation}
\begin{eqnarray}
\hat\mathcal{F}\left( {q_c^2} \right) &=& {F_S}\left( {q_c^2} \right) + {F_V}\left( {q_c^2} \right)\gamma _p^\mu {\gamma _{j\mu }} + {F_T}\left( {q_c^2} \right)\sigma _p^{\mu \nu }{\sigma _{j\mu \nu }} \nonumber \\
&&+ {F_P}\left( {q_c^2} \right)\gamma _p^5\gamma _j^5 + {F_A}\left( {q_c^2} \right)\gamma _p^5\gamma _p^\mu \gamma _j^5{\gamma _{j\mu }} .
\label{Eq5}
\end{eqnarray}
%\end{equation}
%(5)
Here, $\gamma _{p,j}^\mu $ are the proton and nucleon Dirac matrices, ${\gamma ^5} = i{\gamma ^0}{\gamma ^1}{\gamma ^2}{\gamma ^3}$, ${\sigma ^{\mu \nu }} = i\left[ {{\gamma ^\mu },{\gamma ^\nu }} \right]/2$ are the tensor operators; ${F_S}$, ${F_V}$, ${F_T}$, ${F_P}$, and ${F_A}$ are the scalar, vector, tensor, pseudoscalar and pseudovector amplitude components; $q_c^2 =  - t =  - {(p_p^\mu  - p_p^\mu )^2}$ is the invariant momentum transfer. In the $p-N$ c.m. frame, this amplitude can be related to the amplitude (\ref{Eq4}) by equating the corresponding matrix elements:
\begin{equation}
\fl \quad  \bar u_p^{{{\mu '}_1}}\left( {{{{\bf{k'}}}_c}} \right)\bar u_j^{{{\mu '}_2}}\left( { - {{{\bf{k'}}}_c}} \right){\hat \mathcal{F}_j}\left( {q_c^2} \right)u_p^{{\mu _1}}\left( {{{\bf{k}}_c}} \right)u_j^{{\mu _2}}\left( { - {{\bf{k}}_c}} \right) = \chi _p^{{{\mu '}_1} + }\chi _j^{{{\mu '}_2} + }{f_j}\left( {{{\bf{q}}_c}} \right)\chi _p^{{\mu _1}}\chi _j^{{\mu _2}} ,
\label{Eq6}
\end{equation}
%(6)
where ${\chi _{p,j}}$ are the spinors of the proton and nucleon. Note that expressions (\ref{Eq4}) and (\ref{Eq5}) for the $pN$ amplitudes are suitable for scattering by a free nucleon, and the relation (\ref{Eq6}) between them corresponds to m.e. values between the states with positive energies. Thus, the $pN$ amplitude is associated with the results of a phase analysis of the $pN$ scattering, as it is also made in the conventional MDST or in the initial RIA version (IA1; see \cite{McN,Ray2,Ray,Clar,Sak}). However, the amplitude $\hat \mathcal{F}\left( {q_c^2} \right)$ can produce transitions involving states with negative energy, which may be significant for certain virtual transitions, and the relation (\ref{Eq6}), strictly speaking, does not provide the complete information about these effects. However, at sufficiently high energies considered by us, the IA1 variant of RIA turns out to be quite good (see, for example, \cite{Ray,Clar}), which justifies the chosen determination of the $pN$ amplitudes. Further, the covariant $t$-operator is determined using the following relation to the non-covariant m.e.:
%\begin{equation}
\begin{eqnarray}
u_p^{ + {{\mu '}_1}}&&\left( {{{{\bf{k'}}}_c}}\right)u_j^{ + {{\mu '}_2}}\left( { - {{{\bf{k'}}}_c}} \right)\left\langle {{{{\bf{k'}}}_c}} \right|{t^{(j)}}\left| {{{\bf{k}}_c}} \right\rangle u_p^{{\mu _1}}\left( {{{\bf{k}}_c}} \right)u_j^{{\mu _2}}\left( { - {{\bf{k}}_c}} \right) \nonumber \\
&&= 4{\varepsilon _{pc}}\chi _p^{{{\mu '}_1} + }\chi _j^{{{\mu '}_2} + }\left\langle {{{{\bf{k'}}}_c}} \right|{t^{(j)}}{\left| {{{\bf{k}}_c}} \right\rangle _{\rm{nc}}}\chi _p^{{\mu _1}}\chi _j^{{\mu _2}}.
\label{Eq7}
\end{eqnarray}
%\end{equation}
%(7)
This allows us, taking into account the above formulas for the $pN$ amplitudes, to find the covariant definition for the m.e. of $t$-operator on the energy shell, in particular, being valid in the $p-A$ c.m. frame:
\begin{equation}
\left\langle {{\bf{k'}}} \right|{t_j}\left( {{{\bf{r}}_j}} \right)\left| {\bf{k}} \right\rangle  =  - 16\pi {\varepsilon _{pc}}\exp \left( {i{\bf{q}}{{\bf{r}}_j}} \right)\gamma _p^0\gamma _j^0{\hat \mathcal{F}_j}\left( {{q^2}} \right) .
\label{Eq8}
\end{equation}
%(8)
Here we do not write explicitly the nucleon momenta. For the quasifree scattering, the initial and final momenta of the proton and nucleon correspond to the kinematics of the "Breit system", when the nucleon in the nucleus c.m. frame moves with a certain momentum, which provides conditions of the conservation of energy and momentum (see \cite{McN,Ray2,Ray }), and the squares of the momentum transfers in the $p-N$ and $p-A$ c.m. frames coincide: $ - t = q_c^2 = {q^2}$. In the case of a rapid decrease of the m.e. with the ${q^2}$ increase, for the off-shell m.e. we may take also the same approximation (\ref{Eq8}). In the coordinate representation of the m.e. we obtain the following local form:
\begin{equation}
\fl \; \left\langle {{\bf{r'}}} \right|{t_j}\left( {{{\bf{r}}_j}} \right)\left| {\bf{r}} \right\rangle  = \delta ({\bf{r'}} - {\bf{r}}){t_j}\left( {{\bf{r}} - {{\bf{r}}_j}} \right)$,  ${t_j}\left( {\bf{r}} \right) =  - \frac{2}{{{\pi ^2}}}{\varepsilon _{pc}}\gamma _p^0\gamma _j^0\int {{d^3}q\exp ( - i{\bf{qr}})} {\hat \mathcal{F}_j}\left( {{q^2}} \right) .
\label{Eq9}
\end{equation}
%(9)

In the relativistic description of the target nucleus \cite{Wale,Hor,Gam,Ser,Meng,Typ,Lal}, if nucleon correlations are neglected, the nucleus wave function ${\Psi _0}$ has the form of an antisymmetrized product of the one-nucleon bispinor functions $\varphi _j^\alpha  = (u_j^{},w_j^{})$, where ${u_j}$ and ${w_j}$ are the spinors of the upper and lower components (${w_j}$ is small in the standard representation). Neglect of the correlations in the nucleus is equivalent to averaging the operators in (\ref{Eq1}) over ${\Psi _0}$, in particular it yields ${H_t}\left( {{\bf{\hat k}},\left\{ {{{\bf{r}}_j}} \right\}} \right) \to \sqrt {{{{\bf{\hat k}}}^2} + {M^2}} $. Further, for a spin-zero nucleus, the one-nucleon densities of significance are the scalar ${\rho _{Sj}}$ and vector ${\rho _{Vj}}$ ones, which are related to ${u_j}$ and ${w_j}$ as follows:
\begin{equation}
\fl \qquad \quad {\rho _{Sj}}(r) = (\Psi _0^ +  \cdot \gamma _j^0\delta ({\bf{r}} - {{\bf{r}}_j}){\Psi _0}) = \frac{1}{{{N_j}}}\sum\limits_{i = 1}^{{N_j}} {\left[ {{{\left| {u({{\bf{r}}_i})} \right|}^2} - {{\left| {w({{\bf{r}}_i})} \right|}^2}} \right]} ,
\label{Eq10}
\end{equation}
%(10)
\begin{equation}
\fl \qquad \quad {\rho _{Vj}}(r) = (\Psi _0^ +  \cdot \delta ({\bf{r}} - {{\bf{r}}_j}){\Psi _0}) = \frac{1}{{{N_j}}}\sum\limits_{i = 1}^{{N_j}} {\left[ {{{\left| {u({{\bf{r}}_i})} \right|}^2} + {{\left| {w({{\bf{r}}_i})} \right|}^2}} \right]}, \; {N_j} = N, Z.
\label{Eq11}
\end{equation}
%(11)
The tensor density, coming from the m.e. of the form $(\Psi _0^ +  \cdot \gamma _j^0\sigma _j^{0i}{\Psi _0})$, may also appear, however its role is rather insignificant. The difference between the densities ${\rho _{Sj}}$ and ${\rho _{Vj}}$ is relatively small (see, for example, \cite{Ray, Hor,Gam}): it is within 10\% in the inner region of nuclei and practically vanishes in the nucleus surface region. However, it may be expected that effect of this difference can mainly be observed in the spin-orbit part of the $pA-$amplitude and affect spin observables (see \cite{Tjon,Wall3}). Earlier, in constructing the MDES model, we proceeded as closely as possible with the usual MDST approach, where only the baryon densities are used, and thus, we also did not take into account the difference in ${\rho _{Sj}}$ and ${\rho _{Vj}}$. To improve the MDES model, it is necessary to take into account the effect of the difference between the scalar and vector densities, trying to improve the quantitative description of the spin observables.

Let us consider the eikonal expansion in powers of $1/{k_0}$ for the propagator ${G^{( + )}} = {\tilde G^{( + )}}{\gamma ^0} = {\left[ {{\gamma ^0}E - {\bgamma \bf{\hat k}} - m - {\gamma ^0}\sqrt {{{{\bf{\hat k}}}^2} + {M^2}}  + i0} \right]^{ - 1}}$. For this purpose, we choose the direction ${{\bf{e}}_z}\parallel {{\bf{k}}_a}$, where ${{\bf{k}}_a} = ({{\bf{k}}_i} + {{\bf{k}}_f})/2$ is the average momentum, and approximately present the operator ${{\bf{\hat k}}^2}$ as follows: ${{\bf{\hat k}}^2} \approx 2{{\bf{k}}_0}({\bf{\hat k}} - {{\bf{k}}_a}) + k_a^2 + {({\bf{\hat k}} - {{\bf{k}}_a})^2}$, where we denote ${{\bf{k}}_0} \equiv {k_0}{{\bf{e}}_z}$. This expansion yields: ${G^{( + )}} \approx {\varepsilon _t}/({\varepsilon _p} + {\varepsilon _t})\left[ {G_e^{( + )} + \delta G_{ne}^{( + )}} \right]$, where the eikonal propagator and the first-order non-eikonal corrections in the coordinate representation may be presented the following form:
\begin{equation}
\fl \quad \left\langle {{\bf{r'}}} \right|G_e^{( + )}\left( {{{\bf{k}}_{i,f}}} \right)\left| {\bf{r}} \right\rangle  =  - \frac{i}{{2{k_0}}}2m{\Lambda _ + }\left( {{{\bf{k}}_{i,f}}} \right){e^{i{k_a}(z' - z)}}\delta ({\bf{b'}} - {\bf{b}})\theta \left( {z' - z} \right),
\label{Eq12}
\end{equation}
%(12)
\begin{eqnarray}
\fl \quad \left\langle {{\bf{r'}}} \right|\delta {G^{( + )}}\left( {{{\bf{k}}_{i,f}}} \right)\left| {\bf{r}} \right\rangle = \frac{1}{{2{k_0}}}{e^{i{k_a}(z' - z)}}\theta \left( {z' - z} \right){\bgamma }\left[ { \mp i{\bf{q}}/2 + \partial /\partial {\bf{b'}}} \right]\delta ({\bf{b'}} - {\bf{b}}) \nonumber \\
\fl \quad + \frac{1}{{4k_0^2}}2m{\Lambda _ + }\left( { - {{\bf{k}}_{f,i}}} \right)\delta ({\bf{r'}} - {\bf{r}}) + \frac{1}{{4k_0^2}}2m{\Lambda _ + }\left( {{{\bf{k}}_{i,f}}} \right){e^{i{k_a}(z' - z)}}\delta ({\bf{b'}} - {\bf{b}})\theta \left( {z' - z} \right).
\label{Eq13}
\end{eqnarray}
%(13)
Here, ${\Lambda _ + }\left( {{{\bf{k}}_{i,f}}} \right) = ({\gamma ^0}{\varepsilon _p} - {\bgamma }{{\bf{k}}_{i,f}} + m)/(2m)$ is the projection operator onto the positive-energy proton states with two choices of the momentum, ${{\bf{k}}_i}$ or ${{\bf{k}}_f}$, used by us further.

Let us introduce the $p-N$ scattering operator averaged over ${\Psi _0}$ together with the factor of the nucleus recoil ${\varepsilon _t}/({\varepsilon _p} + {\varepsilon _t})$, which, taking in to account that $2{\varepsilon _t}{\varepsilon _{pc}}/(({\varepsilon _p} + {\varepsilon _t})m) \approx k/{k_c}$, has the form:
%\begin{equation}
\begin{eqnarray}
\fl \quad {\hat \tau _j}\left( r \right) &\equiv& \frac{{{\varepsilon _t}}}{{{\varepsilon _p} + {\varepsilon _t}}}{\gamma ^0}(\Psi _0^ +  \cdot {t_j}({\bf{r}} - {{\bf{r}}_j}){\Psi _0}) \nonumber \\
\fl &=& - \frac{{{k_0}}}{{{k_c}}}\frac{{2m}}{{{\pi ^2}}}\int {{d^3}q\;{e^{ - i{\bf{qr}}}}\left[ {Q_S^{(j)}(q){F_{Sj}}\left( {{q^2}} \right) + {\gamma ^0}Q_V^{(j)}(q){F_{Vj}}\left( {{q^2}} \right)} \right]},
\label{Eq14}
\end{eqnarray}
%\end{equation}
%(14)
where $Q_{S,V}^{(j)}(q) = \int {{d^3}r\;} {e^{i{\bf{qr}}}}{\rho _{S,Vj}}(r)$ are the scalar and vector formfactors for the nucleon sort $j$.

The $p-A$ scattering amplitude (\ref{Eq3}) can be written in the form $F\left( {{{\bf{k}}_f},{{\bf{k}}_i}} \right) = \left( {{\chi ^{\mu ' + }} \cdot \hat F({\bf{q}}){\chi ^\mu }} \right)$. For providing a $T$-invariant expression in which the $pN$-interaction operator containing the spin rotating term with ${\bsigma \bf{n}}$ can appear at any of $n$ successive collisions, we represent the amplitude operator $\hat F({\bf{q}})$ in such a form that the propagators (\ref{Eq12}) and (\ref{Eq13}) with momenta ${{\bf{k}}_i}$ and ${{\bf{k}}_f}$ are used in the following symmetrized way:
\begin{eqnarray}
\fl \hat F({\bf{q}})= -\frac{1}{{4\pi }}\int {{d^3}r'{d^3}r\;} {e^{i{k_a}(z - z')}}{e^{i{\bf{q}}({\bf{b'}} + {\bf{b}})/2}}2m{B_ + }L( - {{\bf{k}}_f}) \nonumber \\
\fl \times  \left\{ {\delta (r' - r)\sum\limits_{j = 1}^A {{{\hat \tau }_j}\left( r \right)} } \right.
+ \sum\limits_{n = 2}^A {\frac{1}{n}} \sum\limits_{l = 1}^n {\sum\limits_{{i_1} = 1}^A  \ldots  } \sum\limits_{{i_n} \ne {i_1}...}^A \int {{d^3}{r_2} \ldots } {d^3}{r_{n - 1}}{{\hat \tau }_{{i_1}}}\left( {r'} \right) \nonumber \\
\fl \times \left\langle {{\bf{r'}}} \right|G_e^{( + )}\left( {{\bf{k}}_1^{(l)}} \right) + \delta {G^{( + )}}\left( {{\bf{k}}_1^{(l)}} \right)\left| {{{\bf{r}}_2}} \right\rangle {{\hat \tau }_{{i_2}}}\left( {{r_2}} \right) \left\langle {{{\bf{r}}_2}} \right|G_e^{( + )}\left( {{\bf{k}}_2^{(l)}} \right) + \delta {G^{( + )}}\left( {{\bf{k}}_2^{(l)}} \right)\left| {{{\bf{r}}_3}} \right\rangle  \cdots \nonumber \\
\fl \times {{\hat \tau }_{{i_{n - 1}}}}\left( {{r_{n - 1}}} \right) \left\langle {{{\bf{r}}_{n - 1}}} \right|G_e^{( + )}\left( {{\bf{k}}_{n - 1}^{(l)}} \right) + \delta {G^{( + )}}\left( {{\bf{k}}_{n - 1}^{(l)}} \right)\left| {{{\bf{r}}_n}} \right\rangle \left. {{{\hat \tau }_{{i_n}}}\left( {{r_n}} \right)} \right\}L({{\bf{k}}_i}){B_ + }. \;
\label{Eq15}
\end{eqnarray}
Here, ${B_ + } = (1 + {\gamma ^0})/2$ is the projection operator onto the upper components, and $L\left( {{{\bf{k}}_{i,f}}} \right)$ are the Lorentz boosts; the momenta used in the propagators are ${\bf{k}}_m^{(l)} = {{\bf{k}}_f}$ for $1 \le m < l$, and ${{\bf{k}}_m}(l) = {{\bf{k}}_i}$ for $l \le m \le n - 1$, while $1 \le l \le n$. Retaining in (\ref{Eq15}) only the eikonal propagators (\ref{Eq12})  and performing, as described in \cite{Pil}, the calculations with using the representation for the projection operators in the form ${\Lambda _ + }\left( {{{\bf{k}}_{i,f}}} \right) = L\left( {{{\bf{k}}_{i,f}}} \right){B_ + }L\left( { - {{\bf{k}}_{i,f}}} \right)$, we obtain the eikonal expression for the amplitude of the elastic $p-A$ scattering:
\begin{equation}
{\hat F_e}({\bf{q}}) = \frac{{i{k_0}}}{{2\pi }}\int {{d^2}b\;} {e^{i{\bf{qb}}}}\left[ {{\Omega _A}(b) + i\sin \theta \,\,{{\tilde \Omega }_B}(b){\bsigma \bf{n}}} \right],
\label{Eq16}
\end{equation}
%(16)
where the nucleus profile functions are given by the following formulas:
\begin{eqnarray}
\fl {\Omega _A}(b) &=& \int\limits_0^1 {dx} \left\{ {Z\left[ {{{\bar E}_p}(b) + (1 - \cos \theta ){E'}_s^{(p)}\left( b \right)} \right]} \right.{\left[ {1 - x{{\bar E}_p}(b)} \right]^{Z - 1}}{\left[ {1 - x{{\bar E}_n}(b)} \right]^N}  \nonumber \\
\fl &&+ N\left[ {{{\bar E}_n}(b) + (1 - \cos \theta ){E'}_s^{(n)}\left( b \right)} \right]{\left[ {1 - x{{\bar E}_p}(b)} \right]^Z}\left. {{{\left[ {1 - x{{\bar E}_n}(b)} \right]}^{N - 1}}} \right\} ,
\label{Eq17}
\end{eqnarray}
%(17)
%\begin{equation}
\begin{eqnarray}
\fl {\tilde \Omega _B}(b) &=& \int\limits_0^1 {dx} \left\{ {Z{E'}_s^{(p)}\left( b \right){{\left[ {1 - x{{\bar E}_p}} \right]}^{Z - 1}}} \right.{\left[ {1 - x{{\bar E}_n}} \right]^N} \nonumber \\
\fl &&+ N{E'}_s^{(n)}\left( b \right){\left[ {1 - x{{\bar E}_p}} \right]^Z}\left. {{{\left[ {1 - x{{\bar E}_n}} \right]}^{N - 1}}} \right\} .
\label{Eq18}
\end{eqnarray}
%\end{equation}
%(18)
Here, we have introduced the $E$-functions for the nucleon sort $j$ as follows:
\begin{equation}
\fl \qquad {\bar E_j}\left( b \right) = \frac{1}{\pi }\int\limits_0^\infty  {dqq\;{J_0}(qb)\left[ {mQ_S^{(j)}(q){{\tilde F}_{Sj}}\left( {{q^2}} \right) + {\varepsilon _p}Q_V^{(j)}(q){{\tilde F}_{Vj}}\left( {{q^2}} \right)} \right]} ,
\label{Eq19}
\end{equation}
%(19)
\begin{equation}
\fl \qquad {E'}_s^{(j)}\left( b \right) = \frac{1}{{2\pi }}({\varepsilon _p} - m)\int\limits_0^\infty  {dqq\;{J_0}(qb)\left[ {Q_S^{(j)}(q){{\tilde F}_{Sj}}\left( {{q^2}} \right) - Q_V^{(j)}(q){{\tilde F}_{Vj}}\left( {{q^2}} \right)} \right]} ,
\label{Eq20}
\end{equation}
%(20)
with using the notation ${\tilde F_{S,Vj}}\left( {{q^2}} \right) \equiv 4\pi m{F_{S,Vj}}\left( {{q^2}} \right)/(i{k_c})$.

If these proton and neutron $E$-functions can be taken equal (for example, in the averaged form), then the integration can be performed explicitly and we obtain:
\begin{equation}
\eqalign{
{\Omega _A}(b) = \left[ {1 + (1 - \cos \theta )\frac{{{{E'}_s}\left( b \right)}}{{\bar E(b)}}} \right]\left\{ {1 - {{\left[ {1 - \bar E(b)} \right]}^A}} \right\}, \\
{\tilde \Omega _B}(b) = \frac{{{{E'}_s}\left( b \right)}}{{\bar E(b)}}\left\{ {1 - {{\left[ {1 - \bar E(b)} \right]}^A}} \right\}.}
\label{Eq21}
\end{equation}
%(21)

In determining the invariant $pN$-amplitudes entering in the expressions (\ref{Eq19}) and (\ref{Eq20}), as in \cite{Pil}, we again restrict ourselves to the central and spin-orbital terms ${A_j}\left( q \right)$ and ${C_j}\left( q \right)$ in (\ref{Eq4}), because in \cite{Osl} it was shown that in the calculations by the MDST the contributions coming from the terms in (\ref{Eq4}) containing the nucleon spin are small for sufficiently heavy spin-zero nuclei and may be omitted. As in \cite{Pil}, instead of the relation (\ref{Eq6}) between the amplitudes (\ref{Eq4}) and (\ref{Eq5}), we again use a simplified relationship similarly to how it was done in \cite{Ble2}:
\begin{equation}
\fl \qquad {B_ + }L( - {{\bf{k}}_f})\left[ {{F_{Sj}}\left( {{q^2}} \right) + {\gamma ^0}{F_{Vj}}\left( {{q^2}} \right)} \right]L({{\bf{k}}_i}){B_ + } = \frac{1}{{{{(2m)}^2}}}\left[ {{A_j}(q) + q{C_j}(q){\bsigma \bf{n}}} \right] .
\label{Eq22}
\end{equation}
%(22)
Note that, using this relationship, we neglect the initial and final motions of the nucleon in the nucleus and do not also take into account the Wigner rotation for the spin functions (see \cite{McN,Ray}), which is absent in the strictly forward scattering. It is convenient to write the amplitudes in the $p-N$ c.m. frame as: $A\left( q \right) = i{k_c}/(2\pi ){f_c}(q)$, $qC\left( q \right) = i{k_c}/(2\pi ){f_s}(q)$. Then from (\ref{Eq22}) we can obtain the formulas:
\begin{equation}
\fl \qquad {\tilde F_{Sj}}\left( {{q^2}} \right) = \frac{1}{{2({\varepsilon _p} + m)}}f_c^{(j)}(q) - \frac{i}{{\sqrt {k_0^2 - {q^2}/4} }}\left[ {{\varepsilon _p} - \frac{{{q^2}}}{{4({\varepsilon _p} + m)}}} \right]\frac{1}{q}f_s^{(j)}(q) ,
\label{Eq23}
\end{equation}
%(23)
\begin{equation}
\fl \qquad {\tilde F_{Vj}}\left( {{q^2}} \right) = \frac{1}{{2({\varepsilon _p} + m)}}f_c^{(j)}(q) + \frac{i}{{\sqrt {k_0^2 - {q^2}/4} }}\left[ {m + \frac{{{q^2}}}{{4({\varepsilon _p} + m)}}} \right]\frac{1}{q}f_s^{(j)}(q) .
\label{Eq24}
\end{equation}
%(24)

In order to allow for the electromagnetic effects, the central and spin orbital parts of the elastic $p-A$ scattering amplitude should be written in the following form:
%\begin{equation}
\begin{eqnarray}
\fl \qquad {A_0}(q) &=& {A_C}\left( q \right) + i{k_0}\int\limits_0^\infty  {db} b{J_0}(qb) \left\{ \left( {{e^{i{\chi _0}\left( b \right)}} - {e^{i{\chi _1}\left( b \right)}}} \right) \right. \nonumber \\
\fl && \left. + {e^{i{\chi _1}\left( b \right)}}\left[ {{\Omega _A}\left( b \right) - {\Omega _B}\left( b \right){\chi _{1s}}(b)} \right] \right\} ,
\label{Eq25}
\end{eqnarray}
%\end{equation}
%(25)
%\begin{equation}
\begin{eqnarray}
\fl \qquad {B_0}(q) &=& {B_C}\left( q \right) - i{k_0}\int\limits_0^\infty  {db} b{J_1}(qb)\left\{ \left[ {{e^{i{\chi _0}\left( b \right)}}{\chi _{0s}}\left( b \right) - {e^{i{\chi _1}\left( b \right)}}{\chi _{1s}}\left( b \right)} \right] \right. \nonumber \\
\fl && \left. + {e^{i{\chi _1}\left( b \right)}}\left[ {{\Omega _A}\left( b \right){\chi _{1s}}\left( b \right) + {\Omega _B}\left( b \right)} \right] \right\} ,
\label{Eq26}
\end{eqnarray}
%\end{equation}
%(26)
where the nuclear spin-orbit profile function is related to the function (\ref{Eq18}) as follows: ${\Omega _B}\left( b \right) = (i/{k_0})d{\tilde \Omega _B}(b)/db$. Here, the central and spin-orbit parts of the Coulomb phase for the scattering of two point charges are taken in the eikonal form: ${\chi _0}\left( b \right) = 2\xi \ln\left( {{k_0}b} \right)$, ${\chi _{0s}}\left( b \right) = {{2\xi \kappa } \mathord{\left/ {\vphantom {{2\xi \kappa } b}} \right.  \kern-\nulldelimiterspace} b}$; the central and spin-orbit Coulomb amplitudes given by these phases are as follows:
\begin{equation}
\fl \qquad {A_C}\left( q \right) =  - \frac{{2\xi {k_0}}}{{{q^2}}}\frac{{\Gamma \left( {1 + i\xi } \right)}}{{\Gamma \left( {1 - i\xi } \right)}}\exp\left( { - 2i\xi \ln\frac{q}{{2{k_0}}}} \right)$, ${B_C}\left( q \right) =  - i\kappa q{A_C}\left( q \right) ,
\label{Eq27}
\end{equation}
%(27)
where $\xi  = {{Z{e^2}} \mathord{\left/ {\vphantom {{Z{e^2}} {\hbar v}}} \right. \kern-\nulldelimiterspace} {\hbar v}}$ is the Sommerfeld parameter, and the parameter $\kappa $ characterizes magnitude of the spin-orbit part of the $p–A$ Coulomb interaction. The central part of Coulomb phase of the proton scattering on the volume charge of the nucleus in the eikonal approximation is given by \cite{Ahm}
\begin{equation}
\fl \quad \; {\chi _1}\left( b \right) = {\chi _0}\left( b \right) + 8\pi \xi \int\limits_b^\infty  {dr{r^2}\rho _0^{\left( p \right)}\left( r \right)\left[ {\ln\left( {\frac{{1 + \sqrt {1 - {{{b^2}} \mathord{\left/  {\vphantom {{{b^2}} {{r^2}}}} \right.  \kern-\nulldelimiterspace} {{r^2}}}} }}{{{b \mathord{\left/  {\vphantom {b r}} \right.  \kern-\nulldelimiterspace} r}}}} \right) - \sqrt {1 - {{{b^2}} \mathord{\left/  {\vphantom {{{b^2}} {{r^2}}}} \right.  \kern-\nulldelimiterspace} {{r^2}}}} } \right]} ,
\label{Eq28}
\end{equation}
%(28)
and we take its spin-orbit part in the form, analogous to the macroscopic consideration of the interaction of the proton magnetic moment with the Coulomb field of the nucleus in MDST in \cite{Osl}: ${\chi _{1s}}\left( b \right) = (\kappa {k_0})d{\chi _1}\left( b \right)/db$. However, in contrast to the macroscopic value of $\kappa $ used in \cite{Osl}, we choose it to be $\kappa  = {\kappa _{pp}}{k_c}/{k_0}$, where ${\kappa _{pp}}$ is the corresponding electromagnetic spin-orbit parameter for the elementary $pp$-amplitude used by us and, depending on the concrete phase analysis, is defined as ${\kappa _{pp}} = (3 + 4{\mu _a})/4{m^2}$ or ${\kappa _{pp}} = (3/({E_c} + m) + 2{\mu _a}/m)/2{E_c}$, where ${\mu _a} = 1.79$ is the anomalous magnetic moment of proton. This choice has been made in accordance with our previous calculations by MDST (see \cite{Kup1,Kup2}) where we allowed for this interaction microscopically by including the electromagnetic spin-orbit term in the $pN$-amplitudes. Here, the direct use of this microscopic method is complicated, because it would require a special concern about ensuring numerical convergence of the integrals in (\ref{Eq19}), (\ref{Eq20}), however calculations in MDST show that such a simple trick with redefining the parameter $\kappa $ leads to an indiscernible result.

\section{Results of calculations of $p-A$ scattering observables in MDES model}
The above described MDES approach, built by means of the eikonal expansion on the basis of the Dirac equation and using the relativistic nucleon densities and realistic $pN-$scattering amplitudes determined from phase analysis solutions, has been implemented in an original numerical code developed by us. In the presented here calculations with this code, we have used the nucleon densities obtained by us from the microscopic calculations of the nuclear structure in the RMF approximation basing on the RMF models known from the literature, which are based on the nucleus Lagrangian density of the following form
\begin{eqnarray}
\fl {\cal L} &=& \sum\limits_{j = n,p} {{{\bar \Psi }_j}\left\{ {i{\gamma ^\mu }{\partial _\mu } - {m_N} + {g_\sigma }\sigma  - {g_\omega }{\gamma ^\mu }{\omega _\mu } - \frac{{{g_\rho }}}{2}{\gamma ^\mu }{{\brho }_\mu } \cdot {\btau } - \frac{e}{2}{\gamma ^\mu }(1 - {\tau _3}){A_\mu }} \right\}{\Psi _j}} \nonumber \\
\fl &+& \frac{1}{2}{\partial ^\mu }\sigma {\partial _\mu }\sigma  - \frac{1}{2}m_\sigma ^2{\sigma ^2} - \frac{{{g_2}}}{3}{\sigma ^3} - \frac{{{g_3}}}{4}{\sigma ^4} - \frac{1}{4}{\Omega ^{\mu \nu }}{\Omega _{\mu \nu }} + \frac{1}{2}m_\omega ^2{\omega _\mu }{\omega ^\mu } + \frac{{{g_4}}}{4}{(g_\omega ^2{\omega _\mu }{\omega ^\mu })^2} \nonumber \\
\fl &-& \frac{1}{4}{{\bf{R}}^{\mu \nu }}{{\bf{R}}_{\mu \nu }} + \frac{1}{2}m_\rho ^2{{\brho }_\mu } \cdot {{\brho }^\mu } + \frac{{{{\bar g}_{\omega \rho }}}}{2}g_\omega ^2g_\rho ^2{\omega _\mu }{\omega ^\mu }{{\brho }_\mu } \cdot {{\brho }^\mu } - \frac{1}{4}{F^{\mu \nu }}{F_{\mu \nu }} ,
\label{Eq29}
\end{eqnarray}
%(29)
In (\ref{Eq29}), $\sigma $, ${\omega ^\mu }$, and ${{\brho }^\mu }$ are  the scalar, isoscalar-vector, and isovector-vector meson fields $(\mu  = 0,1,2,3)$; ${A_\mu }$ is the photon field and $e$ is the constant of electromagnetic interaction; ${\Psi _{n,p}}$ are the nucleon fields; ${\btau }$ are the isospin Pauli matrices (${\tau _3} = 1$ is for neutron, and ${\tau _3} =  - 1$ is for proton); ${g_\sigma }$, ${g_\omega }$, and ${g_\rho }$ are the meson–nucleon coupling constants; ${m_\sigma }$, ${m_\omega }$, ${m_\rho }$, and ${m_N}$ are the masses of the mesons and nucleon; ${\Omega _{\mu \nu }}$, ${{\bf{R}}_{\mu \nu }}$, and ${F_{\mu \nu }}$ are the tensors of the fields of vector mesons and electromagnetic field. The Lagrangian density (\ref{Eq29}) includes the nonlinear scalar (the constants ${g_2}$ and ${g_3}$) and vector (the constant ${g_4}$) self interactions, as well as the cross interaction of the meson fields (the constant ${\bar g_{\omega \rho }}$). All coupling constants and some of the meson masses were determined in the literature as adjustable parameters of the model from the best description of properties of finite nuclei and nuclear matter. The antiparticle states are not taken into account in the used RMF models. Because of the stationarity and symmetry requirements, the RMF equations involve only the time components of the four-vectors of the nucleon and electromagnetic currents and vector-meson fields. The set of coupled equations includes the Dirac equations for the spinor nucleon fields, the nonhomogeneous Klein–Gordon equations for the meson fields, and the equation for the Coulomb field. In these calculations of the nuclear structure we have obtained a good agreement of the one-particle spectra and bulk characteristics of nuclei with the corresponding results of other authors in the literature.

In our calculations of observables of the $p-A$ scattering by the MDES model, we have used nucleon densities of the target nuclei obtained with various variants of the relativistic effective $NN$-interaction. Below we present the results of such calculations with densities for the RMF models DD-ME2 \cite{Lal}, IUFSU* \cite{Agr}, and NL3* \cite{Lal2}, which provide the description of the ground state properties of finite nuclei with a fairly good accuracy (see, for example, \cite{Lal}). It is worth mentioning characteristic features of these models: in the Lagrangian density (\ref{Eq29}) for DD-ME2 the constants ${g_2}$=${g_3}$=${g_4}$=${\bar g_{\omega \rho }}$=0, while ${g_\sigma }$, ${g_\omega }$, and ${g_\rho }$ depend on the nucleon density in the nucleus; for IUFSU* all terms in (\ref{Eq29}) are present; for NL3* the constants ${g_4}={\bar g_{\omega \rho }}$=0.

In this approach, using the developed numerical code, we have analyzed the complete set of observables, namely, the differential cross sections, analyzing powers, and spin rotation functions for the elastic scattering of $p+^{40}$Ca and $p+^{208}$Pb at the proton energy 800 MeV. In this analysis, for the $p-N$ scattering amplitudes, being another important input to the model, we used the approximating formulas,
$f_c^{(j)}(q) = \left( {{g_{cj}} + {h_{cj}}{q^2}} \right)\exp\left( { - {a_{cj}}{q^2}} \right)$,
$f_s^{(j)}(q) = \left( {{g_{sj}} + {h_{sj}}{q^2}} \right)\exp\left( { - {a_{sj}}{q^2}} \right)$, whose parameters were found in \cite{Kud}, from the phase analysis solution of \cite{Byst}. At the proton energy 800 MeV their values are as follows:
${g_{cp}} = 2.368+i0.015$ fm$^2$, ${h_{cp}} = 0.159+i0.233$~fm$^4$, ${a_{cp}} = 0.227–i0.033$ fm$^2$,
${g_{sp}} = 0.594–i0.221$ fm$^3$, ${h_{sp}} = 0$, ${a_{sp}} = 0.144+i0.008$~fm$^2$; ${g_{cn}} = 1.913+i0.685$ fm$^2$,
${h_{cn}} = 0.105–i0.107$ fm$^4$, ${a_{cn}} = 0.190–i0.117$ fm$^2$, ${g_{sn}} = 0.435–i0.246$ fm$^3$, ${h_{sn}} = 0$,
${a_{sn}} = 0.153+i0.019$~fm$^2$.

%Figure 1
\begin{figure}[th]
\centerline{\includegraphics[width=6.3cm]{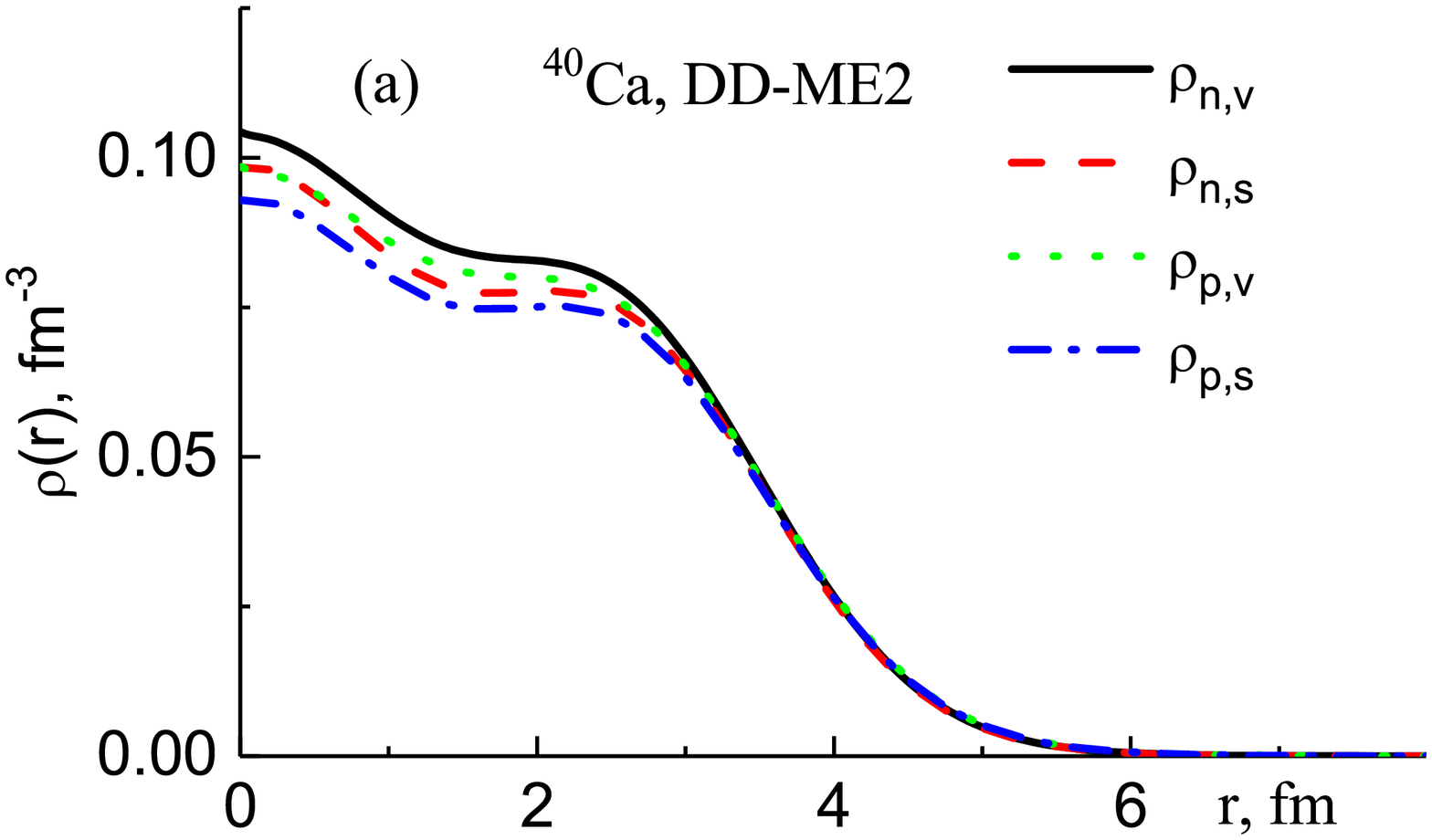}
\;
\includegraphics[width=6.3cm]{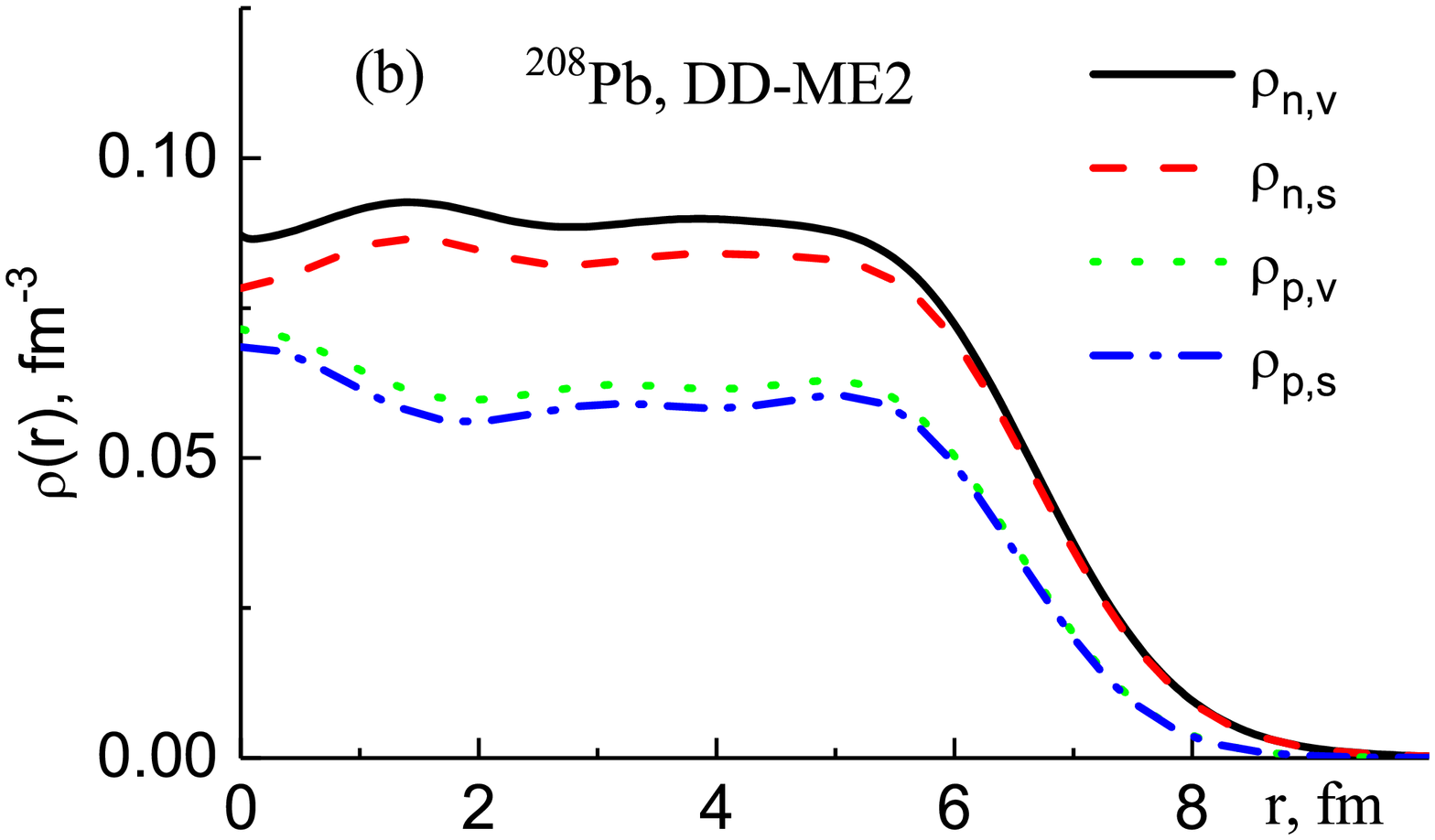}}
\caption{\label{F1}Radial dependences of the vector ${\rho _{nV}}\left( r \right)$, ${\rho _{pV}}\left( r \right)$ and scalar ${\rho _{nS}}\left( r \right)$, ${\rho _{pS}}\left( r \right)$ neutron and proton densities for $^{40}$Ca (a) and $^{208}$Pb (b) nuclei calculated in the RMF model on the basis of the DD-ME2 interaction.}
\end{figure}

In figure~\ref{F1}, we show the results of calculations of the scalar and vector nucleon densities for $^{40}$Ca and $^{208}$Pb nuclei based on the DD-ME2 interaction to illustrate the properties of these densities and the difference between them. As was noted above, a moderate difference between ${\rho _{Sj}}$ and ${\rho _{Vj}}$ (within 10\%) is observed only in the interior of the nuclei; for $^{40}$Ca this difference somewhat exceeds the difference between the neutron and proton densities, and for $^{208}$Pb it is much smaller than the latter one.
%Figure 2
\begin{figure}[th]
\centerline{\includegraphics[width=6.0cm]{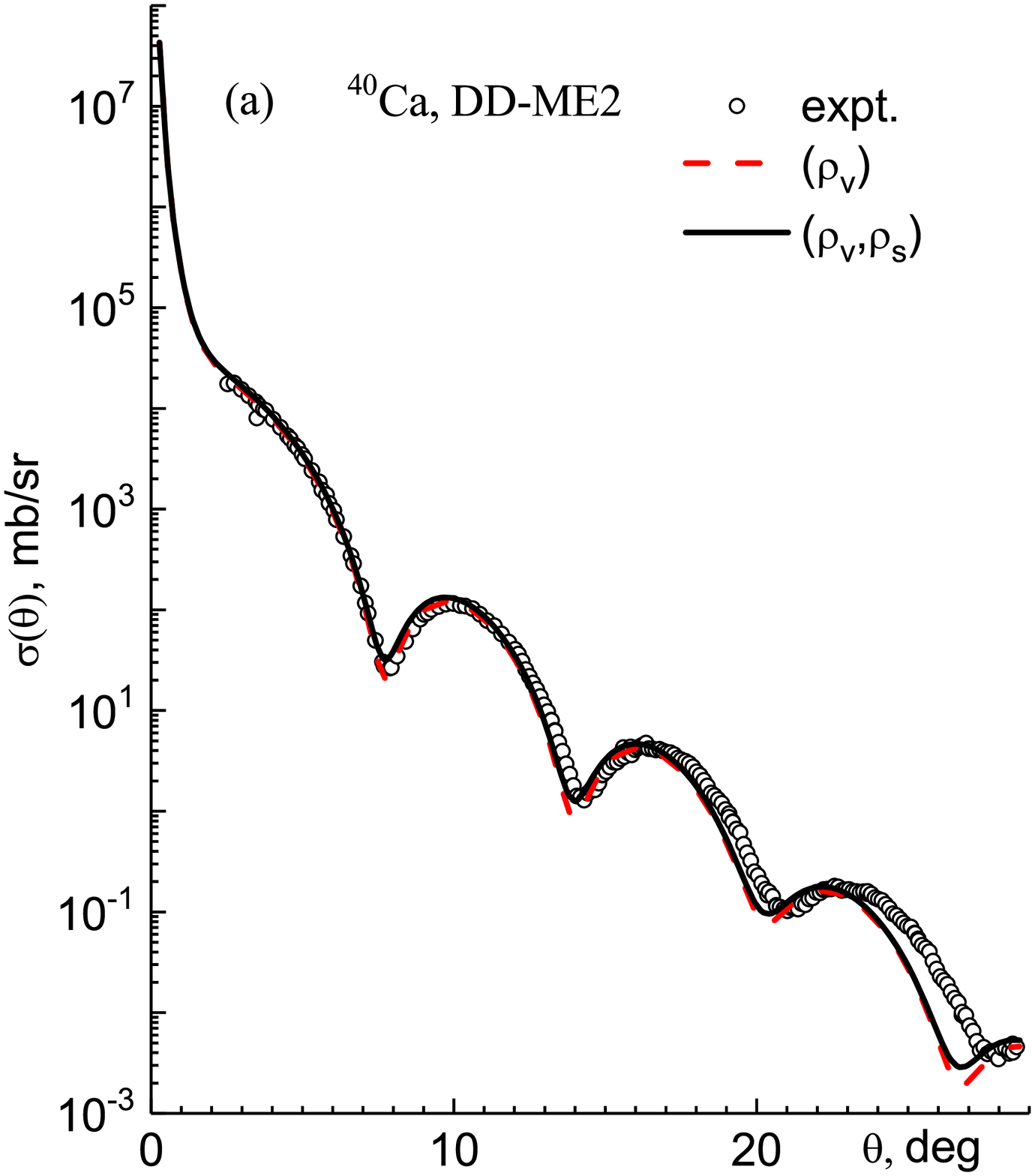}
\;
\includegraphics[width=6.0cm]{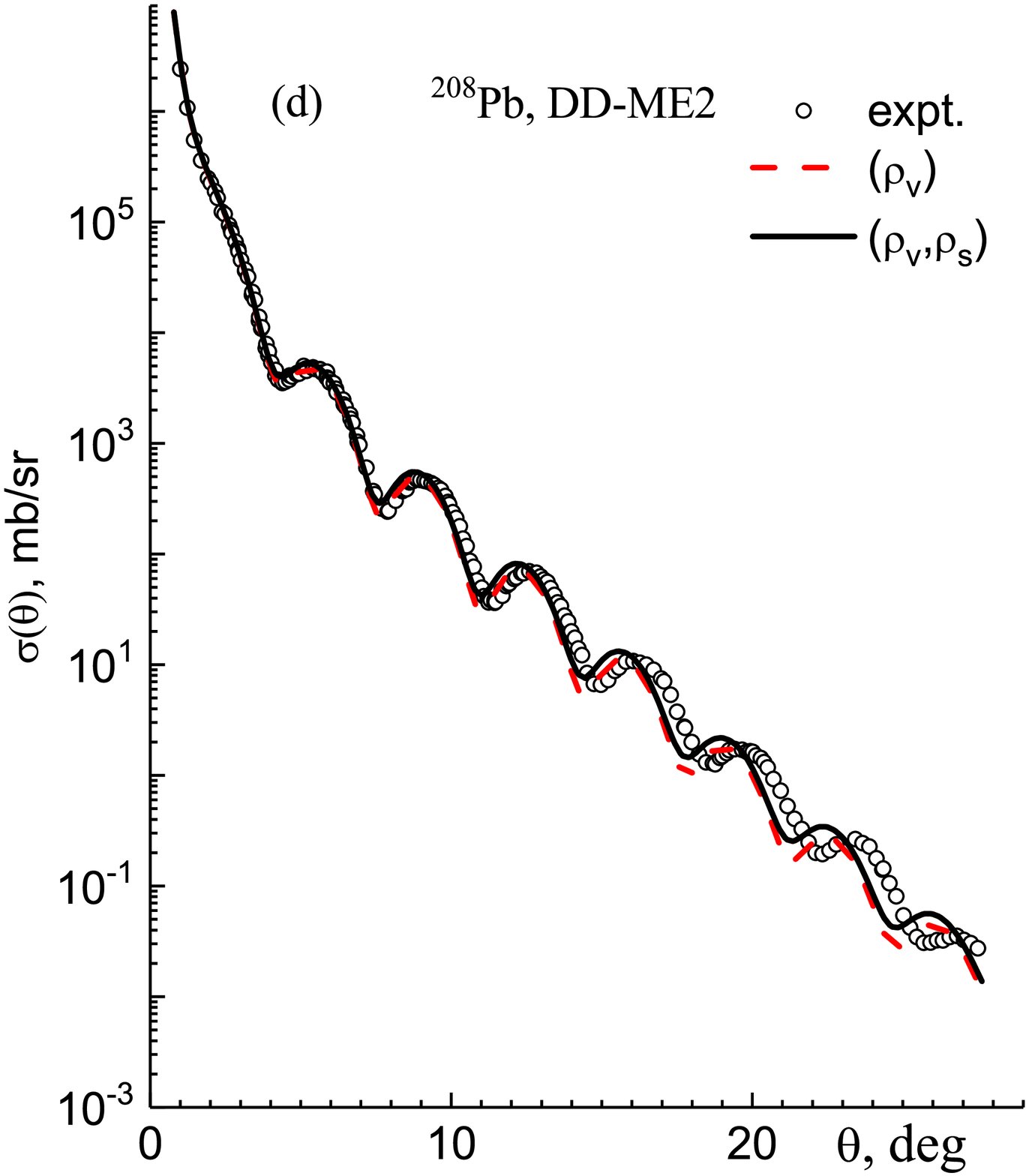}}

\medskip
\centerline{\includegraphics[width=6.0cm]{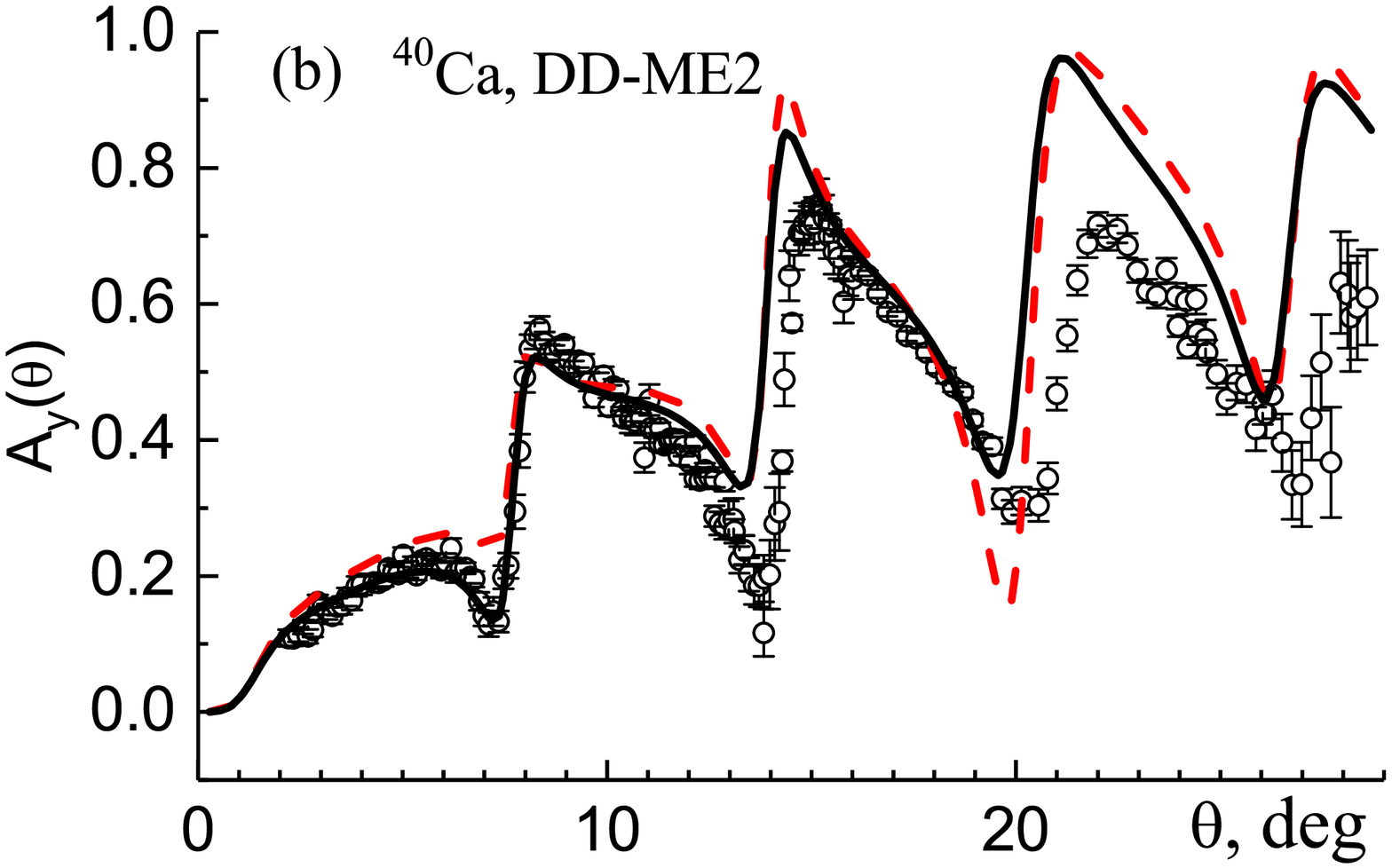}
\;
\includegraphics[width=6.0cm]{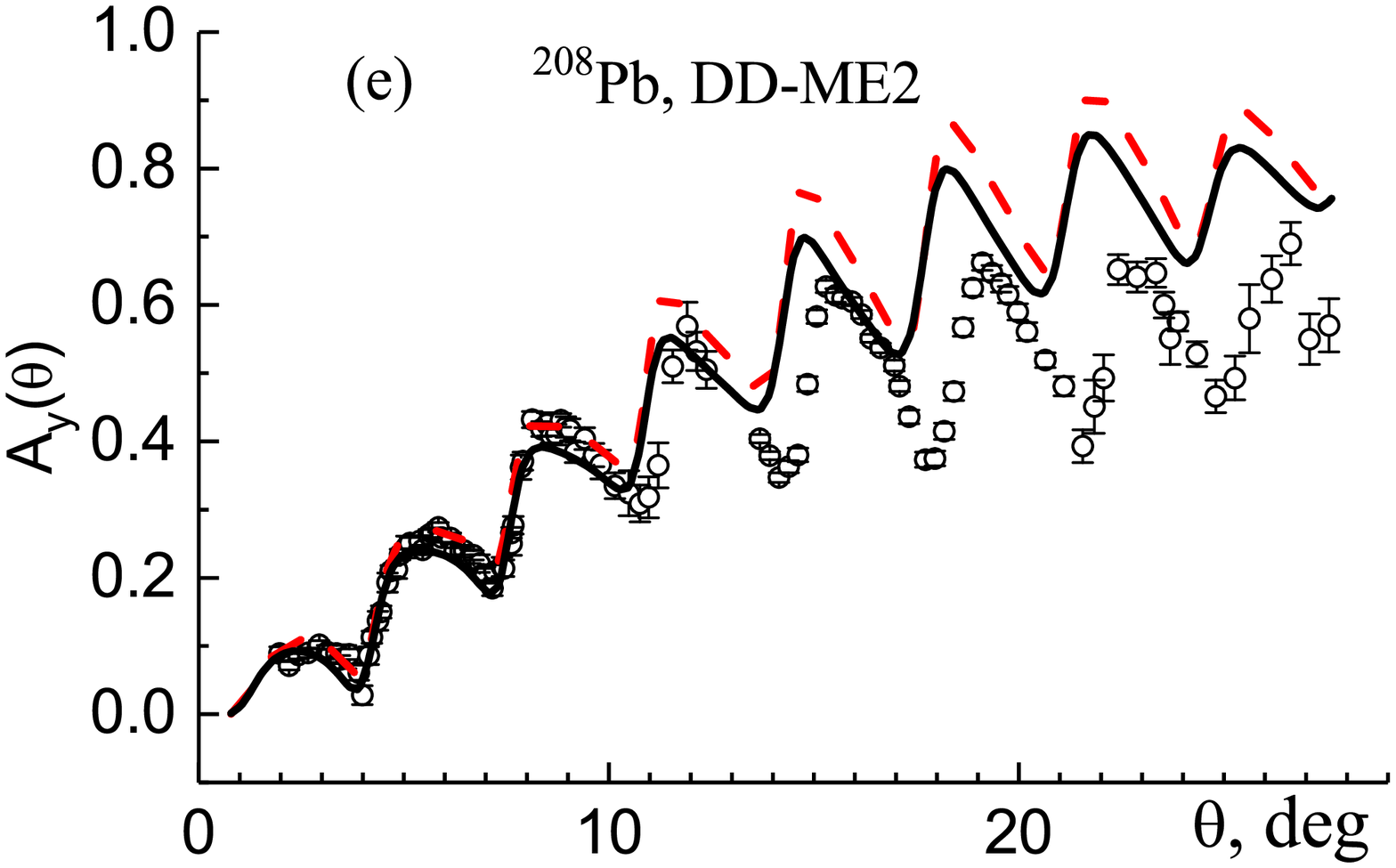}}

\medskip
\centerline{\includegraphics[width=6.0cm]{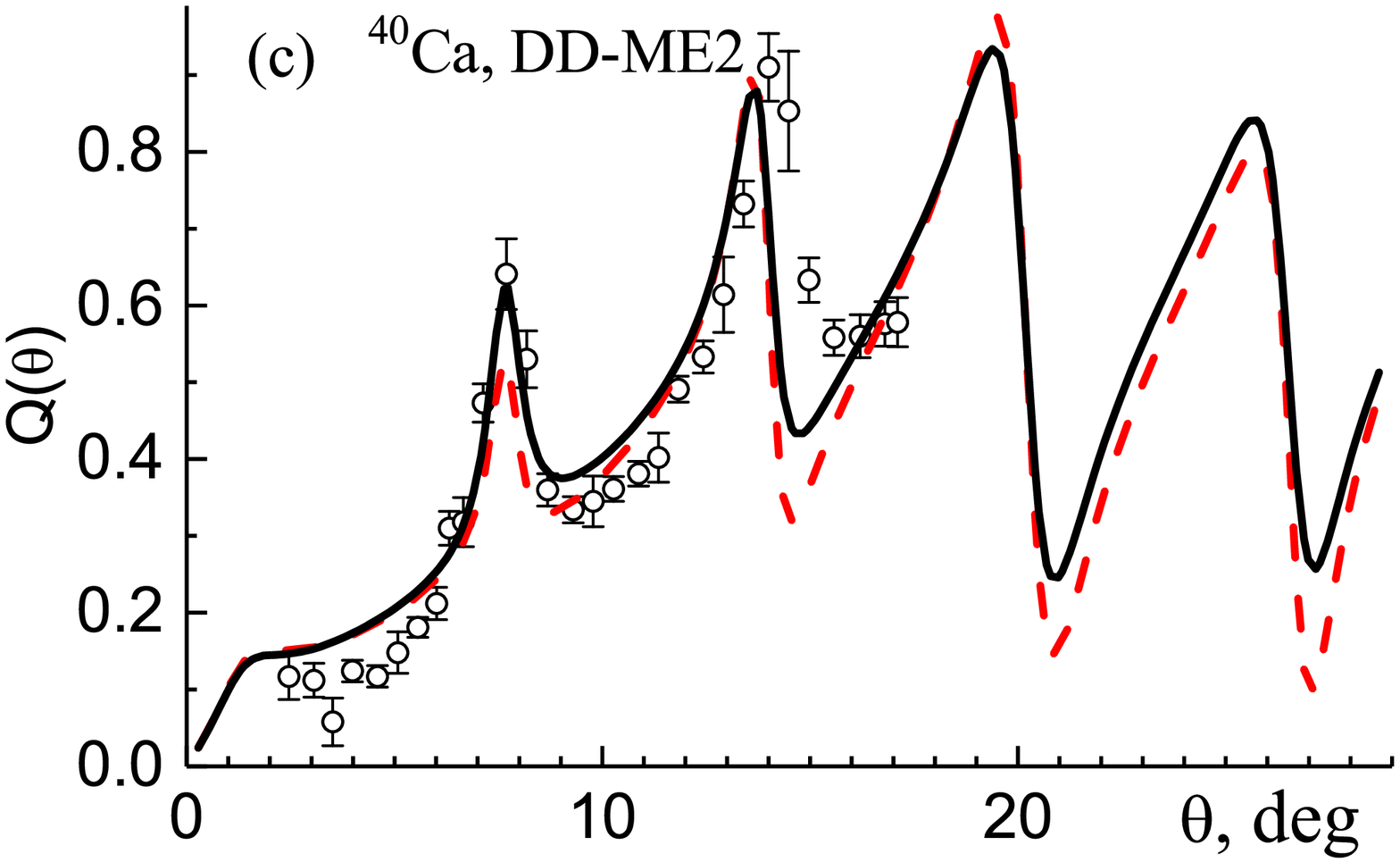}
\;
\includegraphics[width=6.0cm]{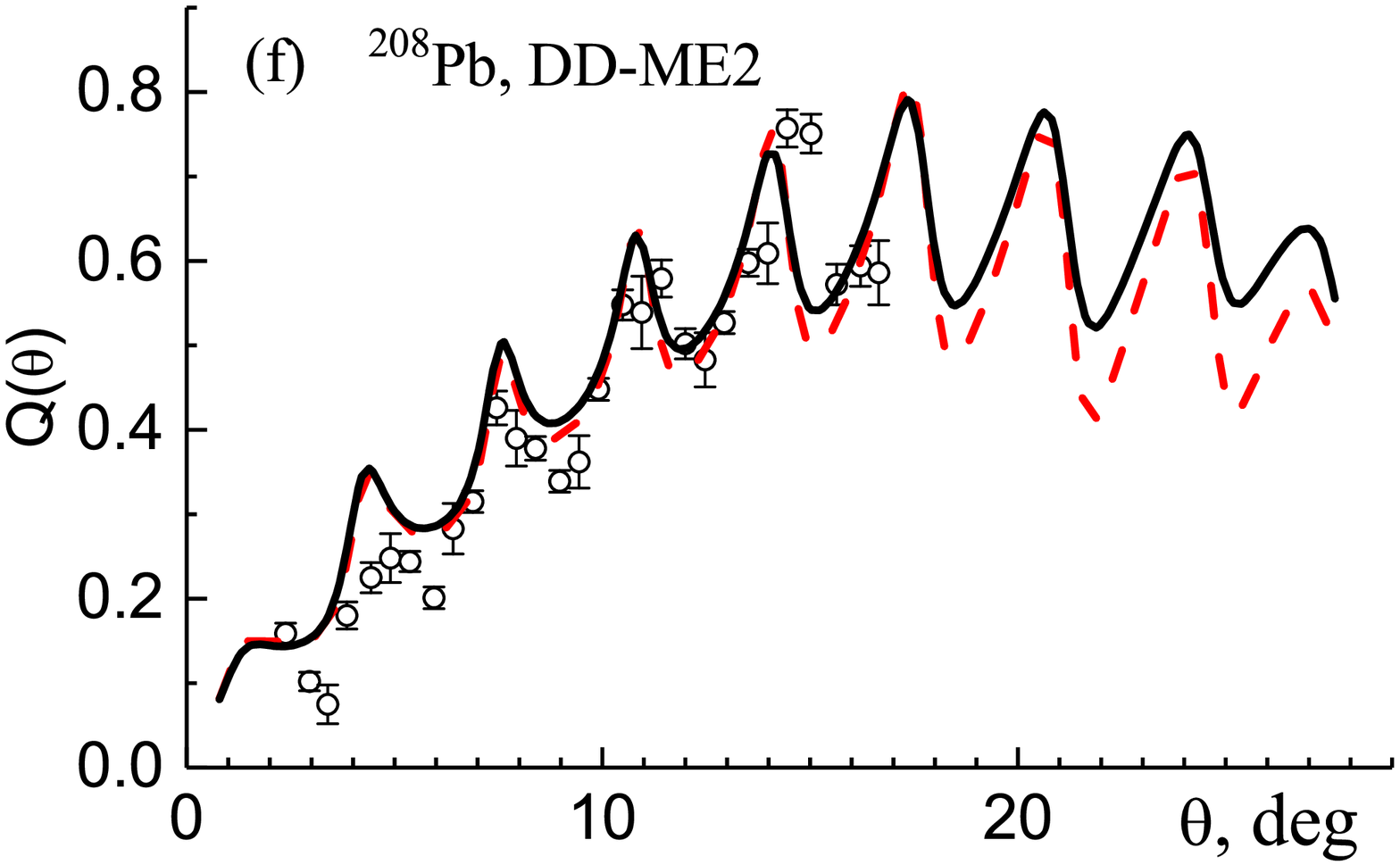}}
\caption{\label{F2}Differential cross sections $\sigma \left( \theta  \right)$, analyzing powers ${A_y}\left( \theta  \right)$, and spin rotation functions $Q\left( \theta  \right)$ for the elastic $p+^{40}$Ca and $p+^{208}$Pb scattering at 800 MeV calculated by the MDES approach with using the nucleon densities obtained in the RMF model with the DD-ME2 interaction: (${\rho _{\rm{V}}}$) is the simplified variant ${\rho _{NS}}\left( r \right) = {\rho _{NV}}\left( r \right)$; (${\rho _{\rm{S}}}$,${\rho _{\rm{V}}}$) is the improved variant of the MDES model with allowing for the difference between the scalar and vector densities. The experimental data are taken from \cite{Ble3,Hof1,Hof2,Ferg}.}
\end{figure}

Further, in figure~\ref{F2}, we present the results of calculations of the differential cross sections, analyzing powers, and spin rotation functions, for the elastic $p+^{40}$Ca and $p+^{208}$Pb scattering at the proton energy 800 MeV, using the relativistic nucleon densities shown in figure~\ref{F1}, performed both in the simplified version of the MDES approach using only the vector (baryon) densities and in the more consistent relativistic version of this model taking into account the difference between the scalar and vector nucleon densities. From figure~\ref{F2} one can clearly see the effect of the difference between these densities on the behavior of the mentioned observables of $pA$ scattering. Although this effect is not too large, however it can be considered to be essential enough. As it was expected, the effect of the distinction between the scalar and vector nucleon densities is more noticeable in the spin observables ${A_y}\left( \theta  \right)$ and $Q\left( \theta  \right)$. It is of interest, that manifestations of this effect are somewhat different for scattering by $^{40}$Ca and $^{208}$Pb nuclei.

For the case of scattering on $^{40}$Ca nuclei, the most significant differences between these two variants of the calculations are observed in the region of diffraction minima of the observables ${A_y}\left( \theta  \right)$ and $Q\left( \theta  \right)$, and it is also worth noting the improvement of the description of the data in the region of the first maximum of ${A_y}\left( \theta  \right)$ when both vector and scalar densities are involved. For the scattering on $^{208}$Pb nuclei, an essential effect of the allowance for the distinction of densities is the lowering in the maxima envelope of the analyzing power ${A_y}\left( \theta  \right)$ observed with the scattering angle increase, although it is not sufficient for a sufficiently good description of the data. Of interest also is that the effect of the difference in these densities is noticeably manifested in the $p-^{208}$Pb scattering differential cross section with increasing the scattering angle.
%Figure 3
\begin{figure}[th]
\centerline{\includegraphics[width=6.0cm]{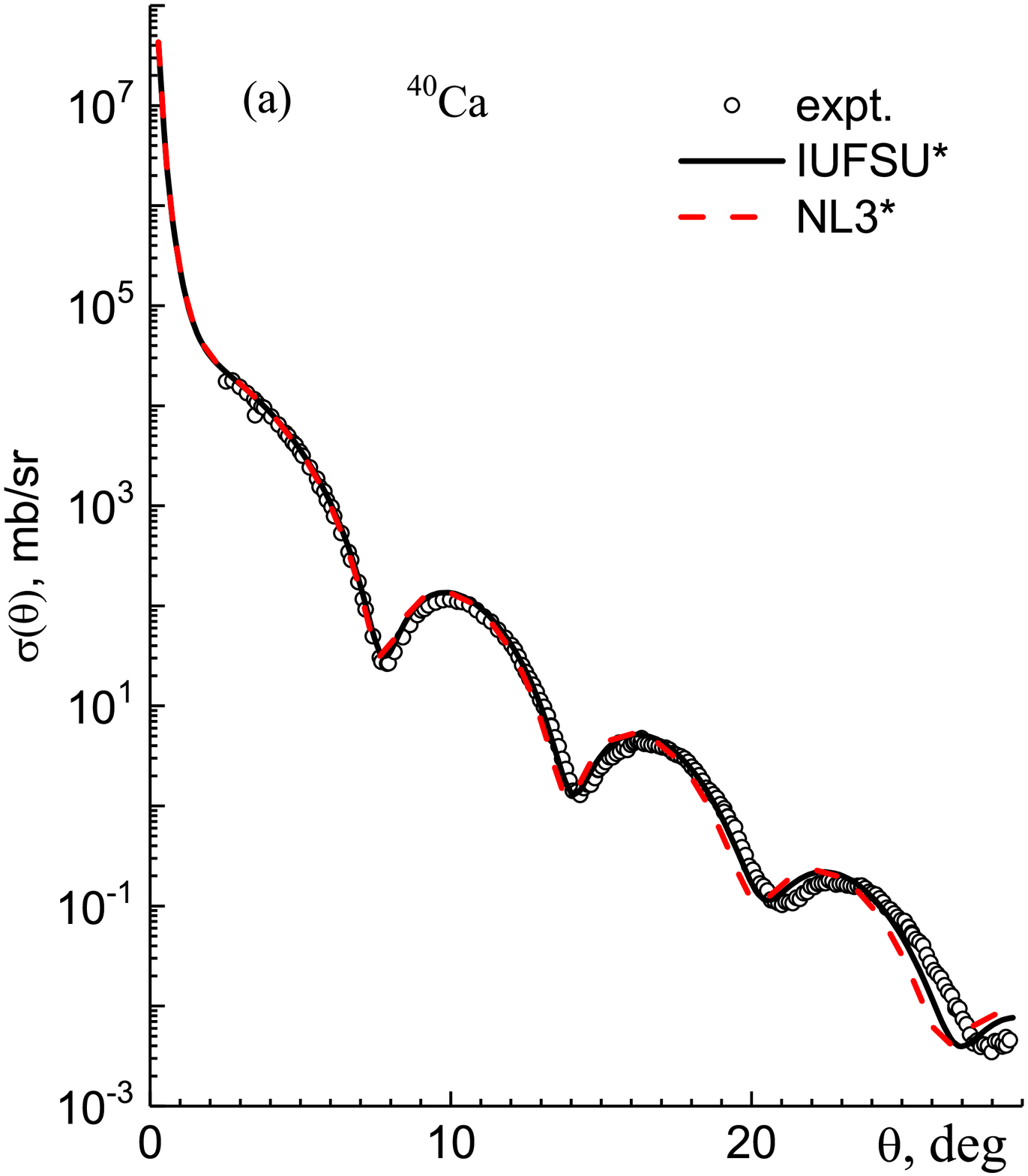}
\;
\includegraphics[width=6.0cm]{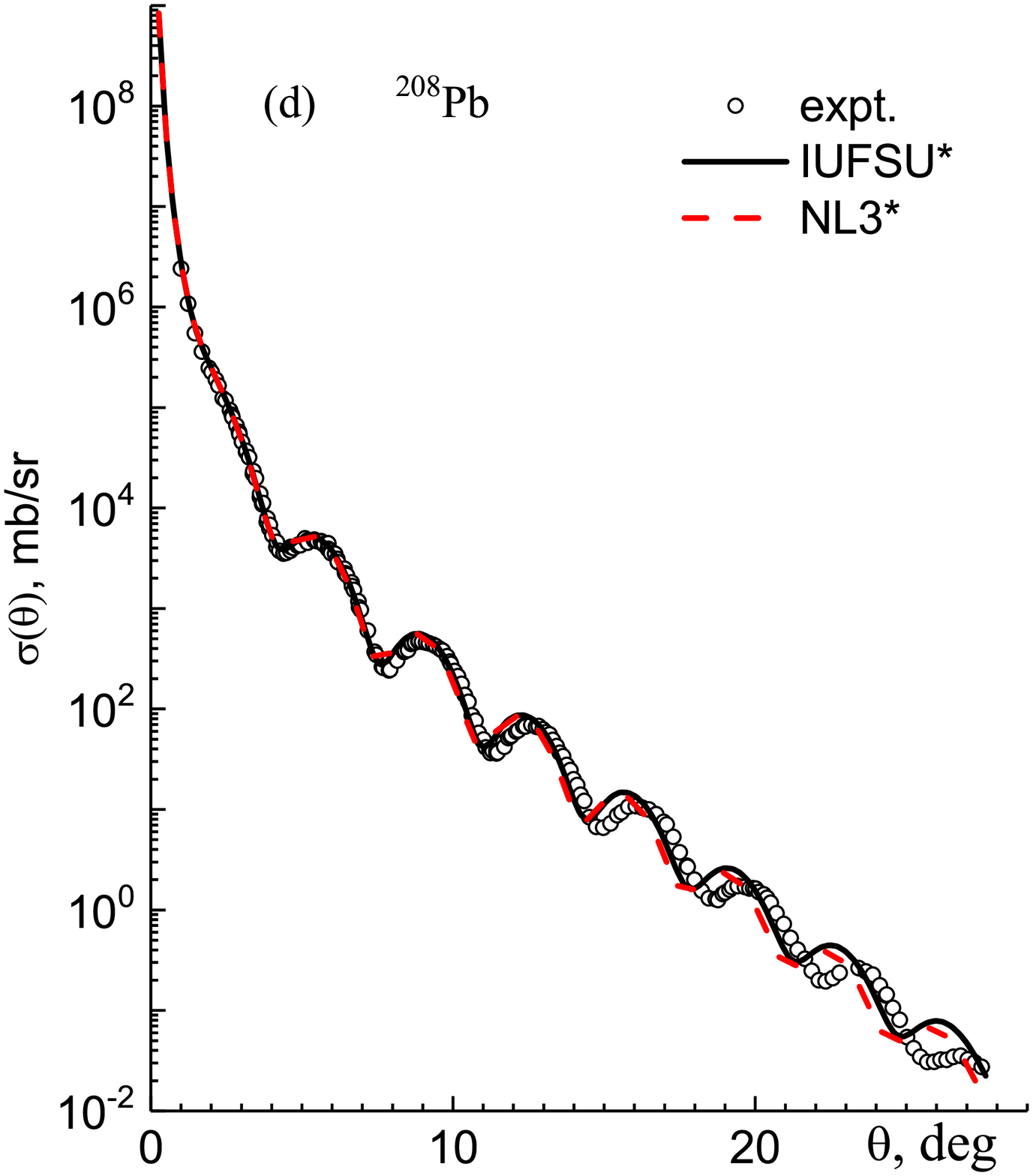}}

\medskip
\centerline{\includegraphics[width=6.0cm]{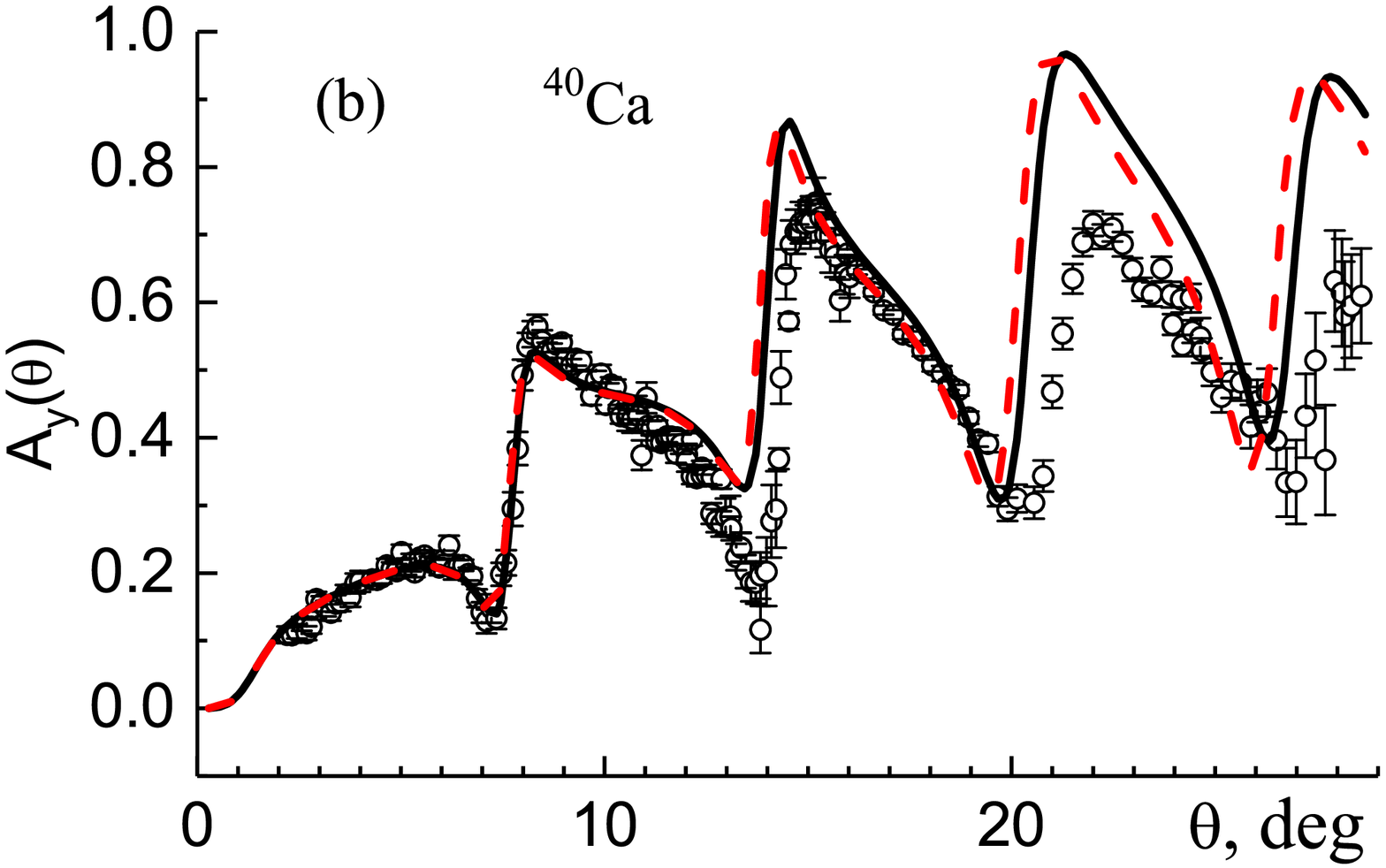}
\;
\includegraphics[width=6.0cm]{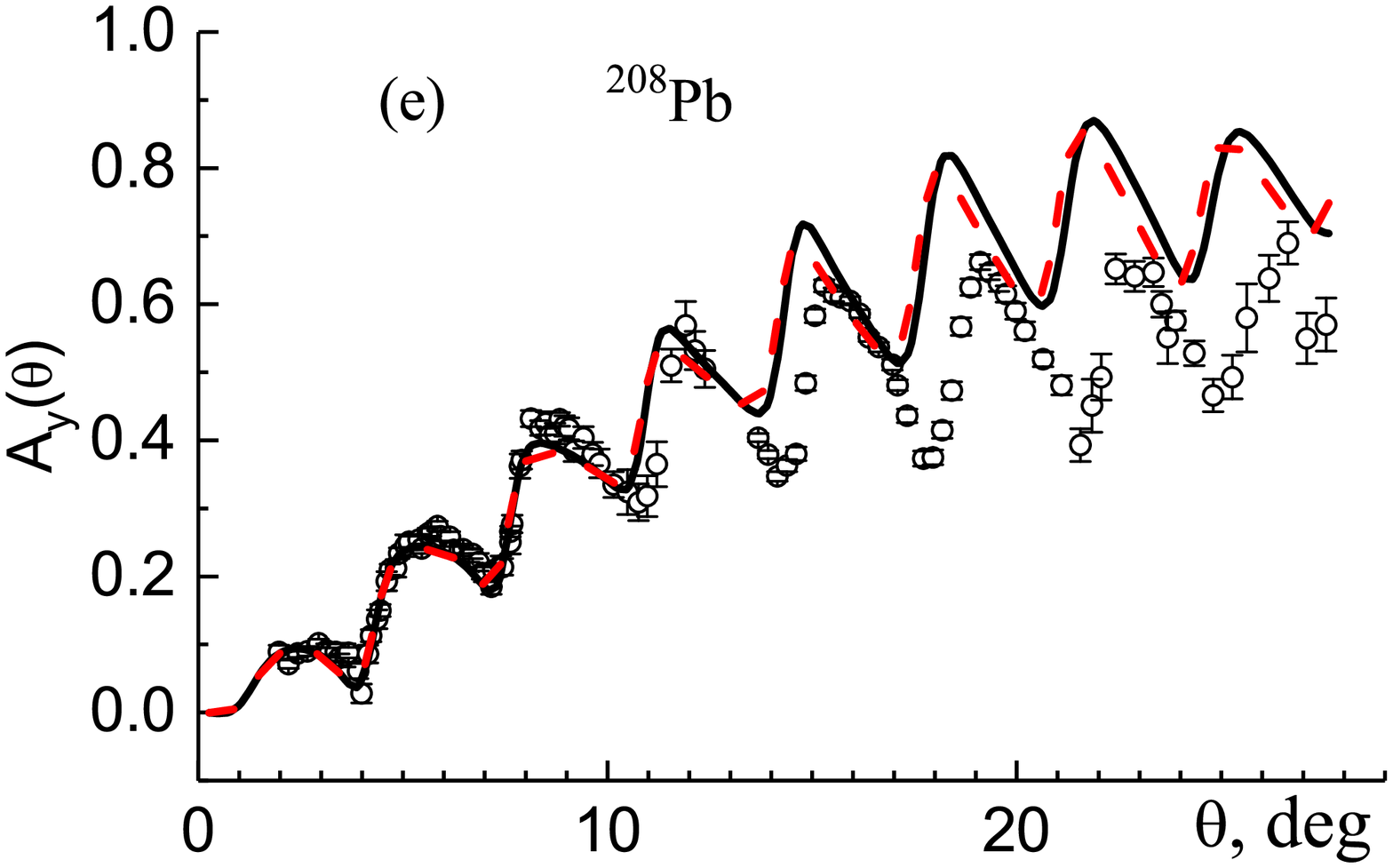}}

\medskip
\centerline{\includegraphics[width=6.0cm]{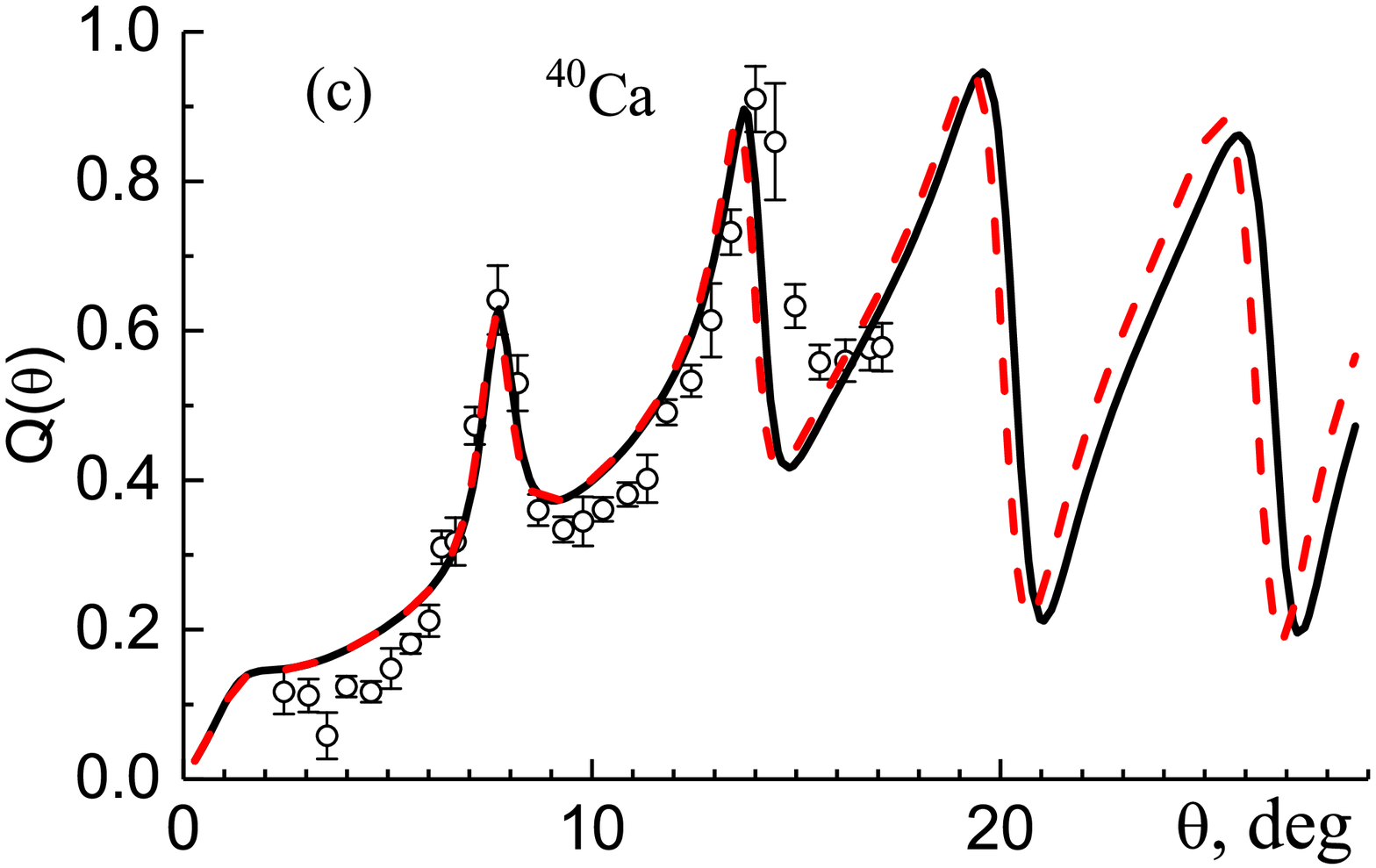}
\;
\includegraphics[width=6.0cm]{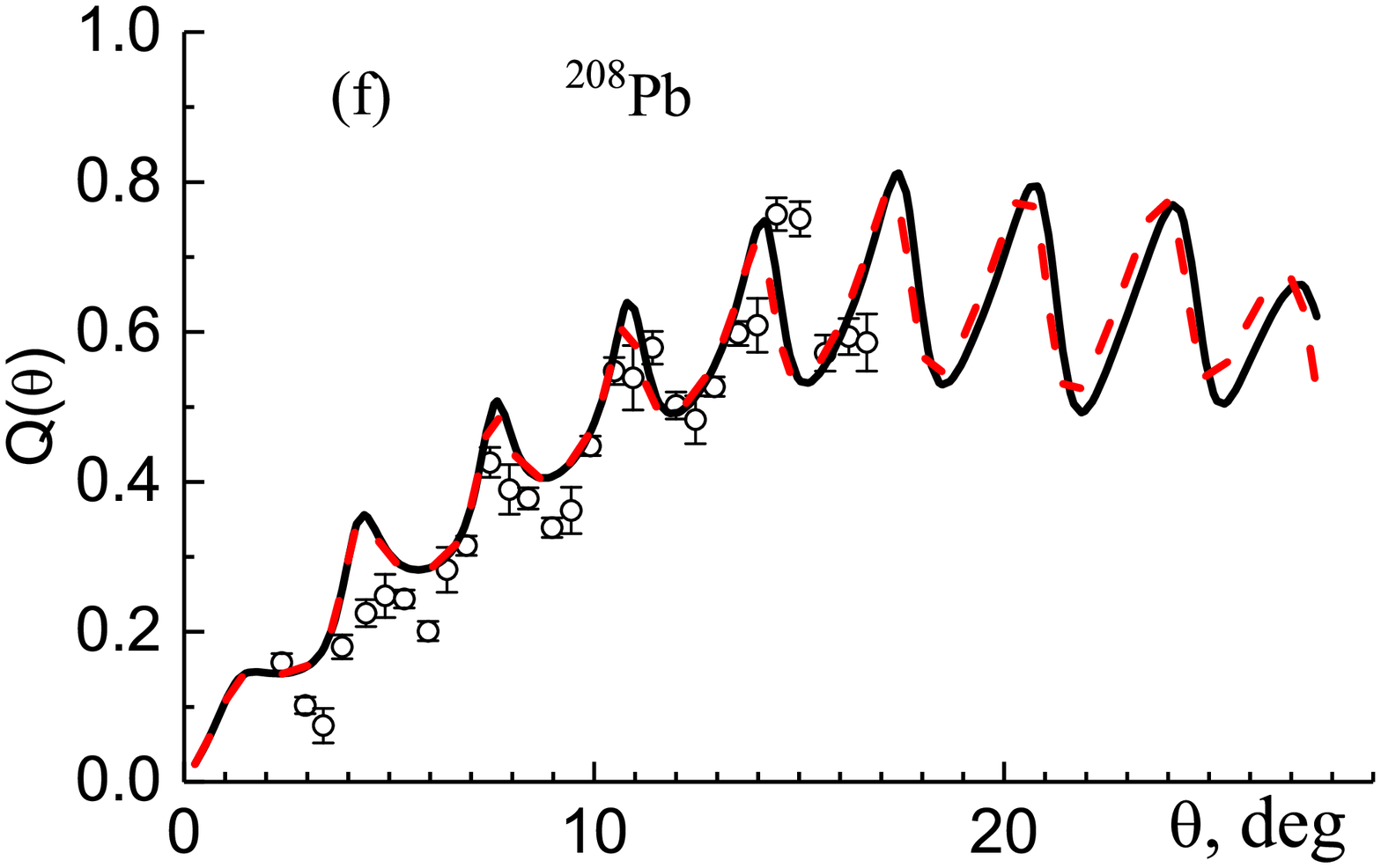}}
\caption{\label{F3}The same, as in figure~\ref{F2}, for the calculations by the improved variant of the MDES model allowing for the difference between the scalar and vector nucleon densities, with using the RMF model for the IUFSU* and NL3* interactions.}
\end{figure}

Figure~\ref{F3} presents the results of calculations of the same scattering observables on the basis of the MDES model with taking into account the difference between scalar and vector densities calculated by the RMF model variants IUFSU* and NL3*. As can be seen from figure~\ref{F3}, the use of the nucleon densities obtained with various considered variants of the relativistic effective $NN$ interaction leads to smaller differences in the calculation results for the observables than the differences of the curves in figure~\ref{F2} due to whether the distinction between scalar and vector densities is taken into account or not.

Assessing the description of the experimental data in figures~\ref{F2} and \ref{F3}, we may say that, in general, the calculated cross sections and spin observables reasonably describe the behavior of the measured quantities, but there are some discrepancies with the data, which are also characteristic of the results of corresponding calculations by the MDST approach (see, \cite{Kup1,Kup2,Pil}). Firstly, a certain shift of the diffraction oscillations toward smaller scattering angles is observed in the calculated observables, which becomes noticeable at larger $\theta $ values. Secondly, although the present version of MDES model, refined as compared to its simpler variant in \cite{Pil}, slightly improves description of the spin observables ${A_y}\left( \theta  \right)$ and $Q\left( \theta  \right)$ (see figure~\ref{F2}), nevertheless it should be noted that there remain certain shortcomings in agreement with the experimental data. This is more conspicuous in the case of the analyzing powers in figures~\ref{F2}b,~\ref{F2}e and \ref{F3}b, \ref{F3}e, where the theoretical curves go somewhat higher than the experimental points with the scattering angle increase for all variants of the calculations, while the calculated spin-rotation functions are in a better agreement with the available data. Therefore, from the presented results a conclusion may be made that we should consider possibilities of further developing the MDES model in order to improve description of the experimental data. In \cite{Pil} we supposed that in the MDES model it may be desirable to consider the effects of intermediate excitations (IE) of target nuclei and noneikonal (NE) corrections analogously to how it was done in the calculations by MDST in \cite{Kup2}. As discussed in \cite{Kup2,Pil}, these corrections in the MDST become essential with the scattering angle increase; the contribution of the IE of nuclei provides a compensation of the above-mentioned shift of the diffraction oscillations of the observables, whereas the NE corrections are mainly manifested in smoothing the diffraction minima. In particular, in the MDST calculations the allowance for the IE and NE corrections improves the description of experimental cross sections at larger angles. Moreover, we should note that the NE corrections in (\ref{Eq13}) contain terms that have no analogs in our MDST calculations and may be expected to affect the spin observables. Another possibility of improving the MDES model may be sought for in studying the role of the omitted terms in the $pN$-amplitude, including the effect of Wigner rotation.

\section{Conclusions}
In this paper, the authors continue to develop the model for a relativistic description of the process of multiple scattering of an incident proton by nucleons of the target nucleus, proceeding by analogy with the well-known multiple diffraction scattering theory by Glauber. For this purpose, a more accurate relativistic formulation has been considered for the model of the multiple Dirac eikonal scattering (MDES) of incident proton on nucleons of the target nucleus, in which expressions for the elastic $p-A$ scattering amplitudes have been derived from the multiple scattering Watson series by means of the eikonal expansion for the propagator of proton free motion, which is performed basing on the Dirac equation with taking into account the recoil of the target nucleus. In this model, we have employed a relativistic description of the target nucleus structure by using realistic nucleon densities calculated on the basis of the relativistic mean field model (RMF). The $pN$-amplitudes, required as another elementary brick of the model, are taken from the results of phase analysis of $pN$ scattering in the literature.

Basing on the MDES approach elaborated by us, the complete set of observables, $\sigma \left( \theta  \right)$, ${A_y}\left( \theta  \right)$, and $Q\left( \theta  \right)$, for the elastic $p+^{40}$Ca and $p+^{208}$Pb scattering at the proton energy of 800 MeV has been calculated. The results of calculations with the use of nucleon densities obtained in different variants of the RMF model, which approved themselves well in the nuclear structure calculations in the literature, have been compared. The effects of taking account of the distinction between the relativistic scalar and vector nucleon densities on the description of the observables of the $p-A$ scattering have been studied. It has been shown that these effects are quite noticeable, especially in the spin observables, and they exceed the differences between the results of calculations with using the nucleon densities obtained in RMF calculations with different modern variants of relativistic effective $NN$ interaction.

Although this more consistent MDES model, correctly allowing for the relativistic scalar and vector nucleon densities, slightly improves description of the spin observables, as compared to its earlier simplified variant, however certain discrepancies with the experimental data indicate that additional refinements of the MDES approach are expedient. Among these, we probably may mention taking into account the contributions to the MDES $p-A$ amplitude, coming from the intermediate excitations of target nuclei and from the noneikonal corrections, as well as studying the effects of the omitted terms in the elementary $pN$-amplitude.

\end{document}